\documentclass{aa}  

\usepackage{graphicx}
\usepackage{txfonts}
\begin{document}

   \title{Probing the galactic cosmic-ray density with current and future $\gamma$-ray instruments }


   \author{G. Peron
          \inst{1}
          \and
         F. Aharonian\inst{1}\fnmsep\inst{2}
          }

   \institute{ Max Plank Institute f\"ur Kernphysik, P.O. Box 103980, D-69029 Heidelberg, Germany  \\
              \email{giada.peron@mpi-hd.mpg.de}
         \and Dublin Institute for Advanced Studies, 10 Burlington Rd, Dublin, D04 C932, Ireland
             \\
             \email{felix.aharonian@mpi-hd.mpg.de}
             }

   \date{Received ; accepted }

 
  \abstract
   {Cosmic Rays (CRs) propagating through dense molecular clouds (MCs) produce gamma-rays which carry direct information about the CR distribution throughout the Galaxy.  Observations of gamma-rays in different energy bands allow exploration of the average CR density in the Galactic Disk, the so-called level of the "CR Sea".  Fermi-LAT observations have demonstrated the method's feasibility based on two dozen MCs in our Galaxy.    However, the potential of Fermi-LAT is limited by the most massive and relatively nearby MCs; thus, the current observations cover only a tiny fraction of the Milky Way.}
   {In this paper, we study the prospects of expanding the CR measurements to very and ultra-high energies and remote parts of the Galaxy with the current and next-generation detectors.   
   }
   {Based on calculations of fluxes expected from MCs, we formulate the requirements to the sensitivity of the post-Fermi-LAT detectors to map GeV-TeV CRs in the Galactic Disk. We also explore the potential of the current and future air-shower and atmospheric Cherenkov telescope arrays for the extension of CR studies to multi-TeV and PeV energy bands.  }
   {We demonstrate that the improvement of the Fermi-LAT sensitivity by a factor of a few would allow a dramatic increase in the number of detectable MCs covering almost the entire Galaxy. The recently completed LHAASO should be able to take the first CR probes at PeV energies in the coming five years or so.    
   }
   {}


   \maketitle
%

\section{Introduction}

The paradigm of Galactic cosmic rays (CRs) assumes that the locally measured CR 
density ($\rho_\odot (\mathrm{1 GeV}) \sim 1 \, \rm eV/cm^{3}$), represents the average 
level of CRs in the Galactic Disk (GD) (see, e.g. \cite{Amato2021}). \ {During their confinement in the GD, CRs mix and lose track of their production sites, creating the so-called "CR sea". The spatial distribution of 
CRs in the Milky Way depends on the distributions of CR sources and  the diffusion coefficient characterizing the CR propagation in GD. It is believed that 
the mixture of CRs caused by diffusion is so effective that one should expect 
uniform distribution of CRs throughout the Galaxy with almost constant level of the "CR sea". However, significant deviations of the density cannot be 
excluded both on small (tens of parsecs) scales because of the concentration of active or recent CR accelerators and on large (kiloparsec) scales due to the spatial variations of the CR diffusion coefficient. 
}

\ {The locally measured CR fluxes give direct information about  the
"CR sea level" only in a single point in the Milky Way. Meanwhile 
the measurements of the "CR sea" level throughout the Galaxy is of paramount importance}. 
Low-energy (MeV/GeV) CRs play a significant role in the regulation of the ionization, chemistry, and the dynamics of the gas and dust, and consequently on the star and planet formation \citep{Padovani2020}. Moreover, CRs might have a non-negligible impact on the habitability of planets around other stars \citep{Atri2014CosmicReview}. At very high energies, the influence of CRs on these processes is less pronounced. However, the information about the distribution of highest energy  CRs in the GD is essential for searching for CR TeVatrons and PeVatrons in the Milky Way. 


Gamma-ray astronomy provides a unique channel for investigating the distribution of CRs far from the Solar System. \ {Of particular interest is the} gamma-ray emission produced at interactions of CRs with the interstellar medium (ISM) \ {which provide straightforward information on the CR content at the location of the interaction}. The observations with the Fermi-Large Area Telescope (LAT) demonstrated the feasibility of \ {this method}: CR densities have been extracted \ {both} from studies of the diffuse gamma-ray emission emission  \citep{Acero2016,Yang2016,Pothast_2018} and from giant molecular clouds (GMCs) \citep{Yang2015,Neronov2017, Aharonian2020, Peron2021}. The latter, being small regions of enhanced gas density, provide localized information on the CR content with accuracy better than 100 pc. 



Fermi-LAT is the only instrument that succeeded in \ {extensively} measuring the $\gamma$-ray flux from "passive", \ {i.e. a cloud without having in the proximity of currently operating CR sources},  GMCs. Yet, the detection is limited to exceptionally massive ($\gtrsim$ 10$^6$ M$_\odot$ ) or nearby (d$\lesssim$ 1 kpc) clouds, \ {if the illuminating CR flux coincides with the local flux of cosmic rays, $J_\odot$}.
The soft power-law spectrum of the "CR sea" ($\alpha_\odot \sim 2.7$)  makes the studies at TeV and higher energies very difficult.  The HAWC Collaboration reported upper flux limits from local molecular clouds \citep{Abeysekara2021} which agree with the above assessment. \ {Meanwhile, the H.E.S.S. Collaboration reported the detection at TeV energies of a GMC located in the galactic plane that shows enhanced emission at GeV energies \citep{Sinha2021SearchSinha}.  } {The advent of new and improved  $\gamma$-ray instruments opens up new possibilities for the exploration of the  sea of galactic cosmic rays in the near future. The Cherenkov Telescope Array (CTA) is designed to reach a sensitivity 10 times better than H.E.S.S. which is promising for detection of at least a few "passive" molecular clouds.  } Even more optimistic assessment can be applied to ultra-high-energy (UHE) gamma-rays thanks to the dramatic improvement of the flux sensitivity above 100~TeV by the Large High Altitude Air Shower Observatory (LHAASO) \citep{Cao2021AnM}. 

Below we discuss the perspectives of detection of gamma-rays from GMCs in high, very-high and ultra-high energy bands. 

\section{Cosmic Ray interaction in molecular clouds}
The inelastic interaction of CRs with the interstellar gas result in  
production of secondary unstable products, first of all $\pi$-mesons, which
decaying, produce gamma-rays, neutrinos and electrons. 
The gamma-ray emissivity  induced in MCs by penetrating CRs depends on (i) the 
energy distribution of CR protons $J(E_p)$, (ii) the density of the ambient hydrogen, and (iii) the content of heavier elements both in the projectile cosmic-rays and in
ambient gas, quantified by the parameter $\xi_N$.
%
The resulting flux at Earth, is given by:

\begin{equation}\label{eq:fluxmc}
F_{\gamma}(E_\gamma) = \xi_N~ \frac{M}{d^{2}}~\int dE_p \frac{d\sigma_{pp\rightarrow \gamma}(E_p,E_\gamma)}{dE_\gamma} J(E_p)\propto A~ \xi_N~ \phi_{\gamma}(E_\gamma), 
\end{equation}
where $A \equiv M_5/d^2_{kpc}$ ($M_5 \equiv M/10^5$ M$_\odot$, $d_{kpc}= d/1$ kpc) is related to the column density of the targeted material and  $d\sigma/dE_{\gamma}$ is the differential gamma-ray 
cross section of proton-proton interactions as calculated by  \cite{Kafexhiu2014}
for the broad interval from  the threshold of pion production 
$\approx 280$~MeV to PeV energies.  

\begin{figure}
	\centering
	\includegraphics[width=1 \linewidth]{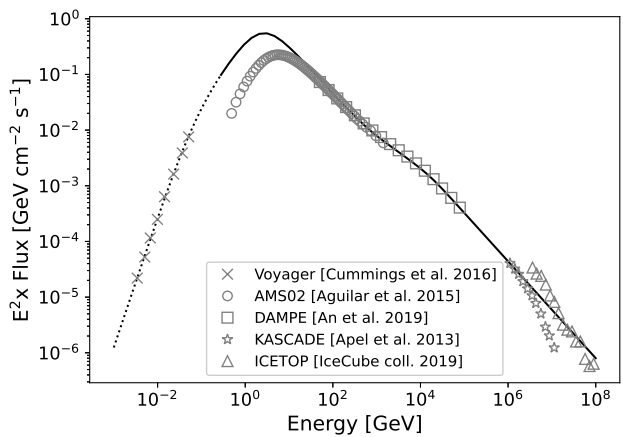}
	\caption{The local spectrum of cosmic ray protons as measured by different experiments (see the figure legend). The black line is the interpolation of the experimental points using the fitting parameters reported by \cite{Vos2015NEW23/24} and \cite{Lipari2020} below and above 100 GeV, respectively  The dotted part represents protons below the energy threshold of 280 MeV, which do not participate in the $\pi$-meson production.}
	\label{fig:proton}
\end{figure}

The local spectrum of protons, $J_\odot$, has been measured with great precision in the Earth's vicinity (e.g. AMS \citep{Aguilar2015}, DAMPE \citep{Amenomori2021}),  
from Earth (e.g. KASCADE \citep{Apel2013}, Icetop \citep{Aartsen2019}), and now  as well outside the Solar System \citep{Stone2019},  where the CR spectrum does not suffer solar modulation. 
Recent measurements revealed that the CR proton spectrum doesn't have a single power law shape; the spectral index $\Gamma$ changes from 2.8 
below $\sim$ 700 GeV to 2.6 up to 15~TeV and steepens again ($\Gamma \approx 2.85$ at higher energies \citep{Lipari2020}. Above 1 PeV, CR measurements are provided by ground-based detectors. The interpretation of indirect measurements depends on the interaction models, therefore they remain controversial. Moreover, the lack of any measurements between 100~TeV and 1~PeV introduces additional 
uncertainties both in the proton spectrum and composition of heavier nuclei. This issue is comprehensively discussed by \cite{Lipari2020}. Fig.\ref{fig:proton} shows 
the recent local measurements of the CR proton fluxes. \ {The black line has been obtained using the fitting parameters derived by \cite{Lipari2020} above 100 GeV, while below 100 GeV, to account for the solar modulation, the curve matches the data of Voyager to the data of AMS as done in \cite{Vos2015NEW23/24}}
  
At low energies, for the standard  compositions  of the  interstellar medium  and CRs, the contribution of  nuclei to the gamma-ray production is comparable to the contribution from the $pp$-interactions, namely $\xi_N \approx 1.8$ \citep{Mori2009}. At higher energies, especially around the 
{\it knee} at PeV energies,  CRs become "heavier"; 
consequently the nuclear  enhancement factor $\xi_N$ increases with energy. The significant uncertainty in the CR composition in the knee region introduces non-negligible uncertainty in  $\xi_N$. The calculations based on the 
available CR data show that $\xi_N$ progressively increases from 1.8 at 10~GeV to $\simeq 2.6$ at 1~PeV. 

The third parameter that determines the cloud's flux is 
$A=M_5/d_{\rm kpc}^2$ which is the measure of the column 
density of the gas embedded in the cloud. Indeed, given that 
$M=N_{col}\theta d^2$, we have $A\propto N_{col}\theta$ where 
$\theta$ is the angular area of the considered region. 
It can be presented in the form:
\begin{equation} A= 8 \times 10^{-20} \sum_{l,b} \bigg( \frac{N_{col}(l,b)}{ \mathrm{cm^{-2}} } \bigg) dl db \bigg(\frac{\pi}{180}\bigg)^2 
\end{equation}
where $dl$ and $db$ are the pixel size of the gas tracer map.
Remarkably,  $A$ is independent of uncertainties both 
in the mass of the cloud and the distance. The only relevant 
uncertainty is  related to the column density and comes from the tracers of molecular gas. The most commonly 
employed tracers are the $^{12}$CO(J=1$\rightarrow$0) line that brings an uncertainty of the order of 30\% \citep{Bolatto2013} in the gas density  and the 
dust opacity with an uncertainty that 
amounts to $\sim$ 20\% \citep{Ade2011}.  

\section{Molecular Clouds in the Milky Way}
\begin{figure}
\centering
\includegraphics[width=1\linewidth]{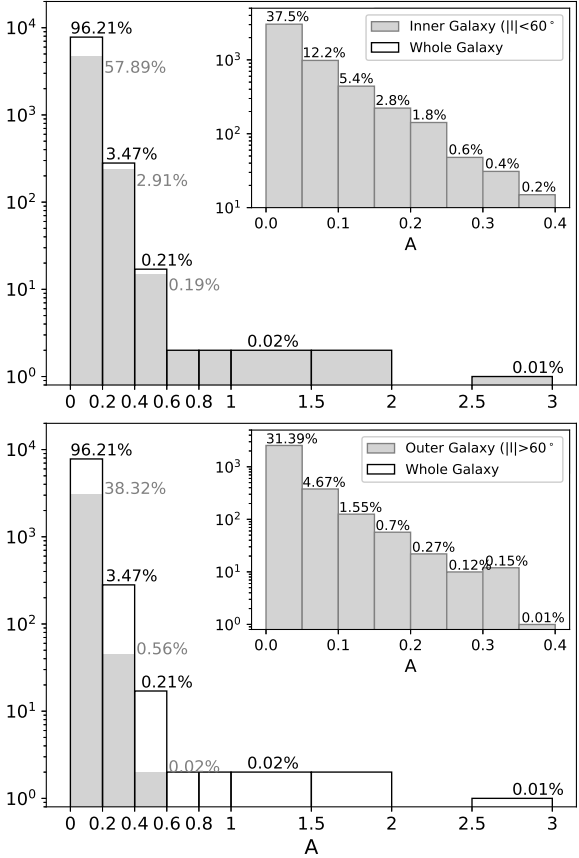}
\caption{Histogram of the $A$ parameter of molecular clouds from the MD16-catalog. The black-contoured bars include all the clouds in the catalog, while the grey bars correspond, in the upper panel, to MCs in the inner Galaxy ($|l|<60^\circ$), and to MCs in the outer Galaxy ($|l|>60^\circ$), in the lower panel. The number fraction with respect to the total is also reported as percentages. The inset panels zoom the parameter range below 0.4 determined  as the detection threshold of MCs  by Fermi-LAT  \citep{Aharonian2020}.} 
\label{fig:Ahist}
\end{figure}

From the recent analysis of the all-Galaxy CO survey of Dame, Hartmann and Thaddeus \citep{Dame2000}, \cite{Miville-Deschenes2016} identified more than 8000 MCs distributed all throughout the galactic plane. When inspecting the clouds of \cite{Miville-Deschenes2016}, hereafter MD16, we see that most of the clouds have a low A parameter  (see Fig \ref{fig:Ahist}),  below 0.4, which was determined in \cite{Aharonian2020} to be a safe threshold  for  spectral measurements of clouds of different extensions,  located both in the inner and outer parts of the Galaxy. \ {These considerations were based on the assumption that the level of CRs that illuminates the cloud is coincident with the local level of CR, $J_\odot$, which was taken as a reference value.}
Among the M16 catalog clouds, less than 1\% is above the detection threshold of Fermi-LAT;  the fraction is even lower when considering only the inner ($|l|<60^\circ$) Galactic regions ($\sim$0.3\%).  For the outer Galaxy, the threshold can be lowered by a factor of 2. However this part of the Galaxy hosts less dense clouds, with $A$  parameters in most cases lower than 0.6 and which overcomes the detection threshold only for the $\sim$ 1\% of the cases, even when lowering the threshold to $A=$0.2. This means that most of the  "CR Sea"  cannot be  explored by Fermi-LAT.  In particular, $\sim$15\% of the molecular clouds, corresponding to more than 1000 MCs, have an $A$ factor between 0.1 and 0.4, just below the Fermi-LAT detection threshold.

In addition to emissivity, source confusion affects the detectability of clouds in $\gamma$-rays. Confusion can arise both due to the proximity of known $\gamma$-ray sources, and due to other clouds located on the same line of sight.

\subsection{Overlaps with other gamma ray sources}
We included into consideration all reported GeV and TeV $\gamma$-ray sources from the 4FGL (Fermi-LAT; \cite{TheFermi-LATcollaboration2019}), HGPS (HESS Galactic Plane Survey, \cite{Abdalla2018}) and the 3HWC (HAWC; \cite{Albert20203HWC:Sources} ) catalogs which lie  within the radius of 1.1 $\theta$, where $\theta$ is the angular size of the cloud:
\begin{itemize}

	\item 75 \% of clouds do not have an overlapping source  
	\item 3 \% of clouds have at least one overlapping known source 
	\item 22 \% of clouds have only unidentified overlapping sources 
	\item 63\% of the clouds do not have nearby sources within  0.5$^\circ$.
	
\end{itemize}
Clouds without nearby sources are ideal to test the "CR Sea", even though this does not exclude possible contributions of yet unresolved gamma-ray sources. 

\subsection{Fraction of gas on the line of sight}
Differently from the smoothly distributed atomic gas,   the molecular component of the interstellar medium (ISM)  is clumpy and mostly concentrated in dense clouds. \cite{Miville-Deschenes2016} pointed out that the line of sight column densities  in most  ($\approx$60\%) directions 
are contributed by three or fewer molecular clouds;  in the 20\% of cases, the column density is dominated by a single cloud. 
Following the approach  proposed by \cite{Peron2021}, one can derive a relation between the maximum fraction of back- and fore-ground gas ($X$) which can be on the line of sight of a cloud and the level of excess ($N$) with respect to the local $\gamma$-ray emissivity ($\phi_\gamma(J_\odot)$), which can be detected:
\begin{equation}
X < \frac{0.7N}{0.7N +1.3}
\end{equation}

For example, if a cloud has an emissivity larger than the nominal value, by a factor of $N=4$, it would be detected if the fraction of background gas is $X<0.68$ or, in other words, if the fraction of column density belonging to the cloud is at least $32\%$.  For detection of the local CR Sea in molecular clouds ($N=1$), the fraction of gas in the cloud has to be at least  65\%. This guarantees the distinction of the cloud above the background gas, even without subtracting the contribution of the latter. Otherwise, the flux of the cloud, even if enhanced, would be masked by the $\gamma$-ray flux of the back- and foreground gas. To avoid this, it is necessary to model the back- and fore-ground gas as a separate source, as done for example in \cite{Aharonian2020}. This approach, however, is subject to large uncertainties of the CO and HI measurements, which are the only tracers that can be used for 3-dimensional decomposition. \ { Notice nevertheless that, even without a 3-dimensional decomposition, measuring a flux similar to $J_\odot$ coming from a column of gas, is a strong indication that the entire column is emitting at a similar level as the local CR sea. The local flux can be considered a minimum level, as no lower flux has been recorded so far, except for the outermost part of the Galaxy, which are far from the highest concentration of Supernova Remnants (SNRs) and Pulsar Wind Nebulae (PWNe). }

\begin{figure}
\includegraphics[width=1\linewidth]{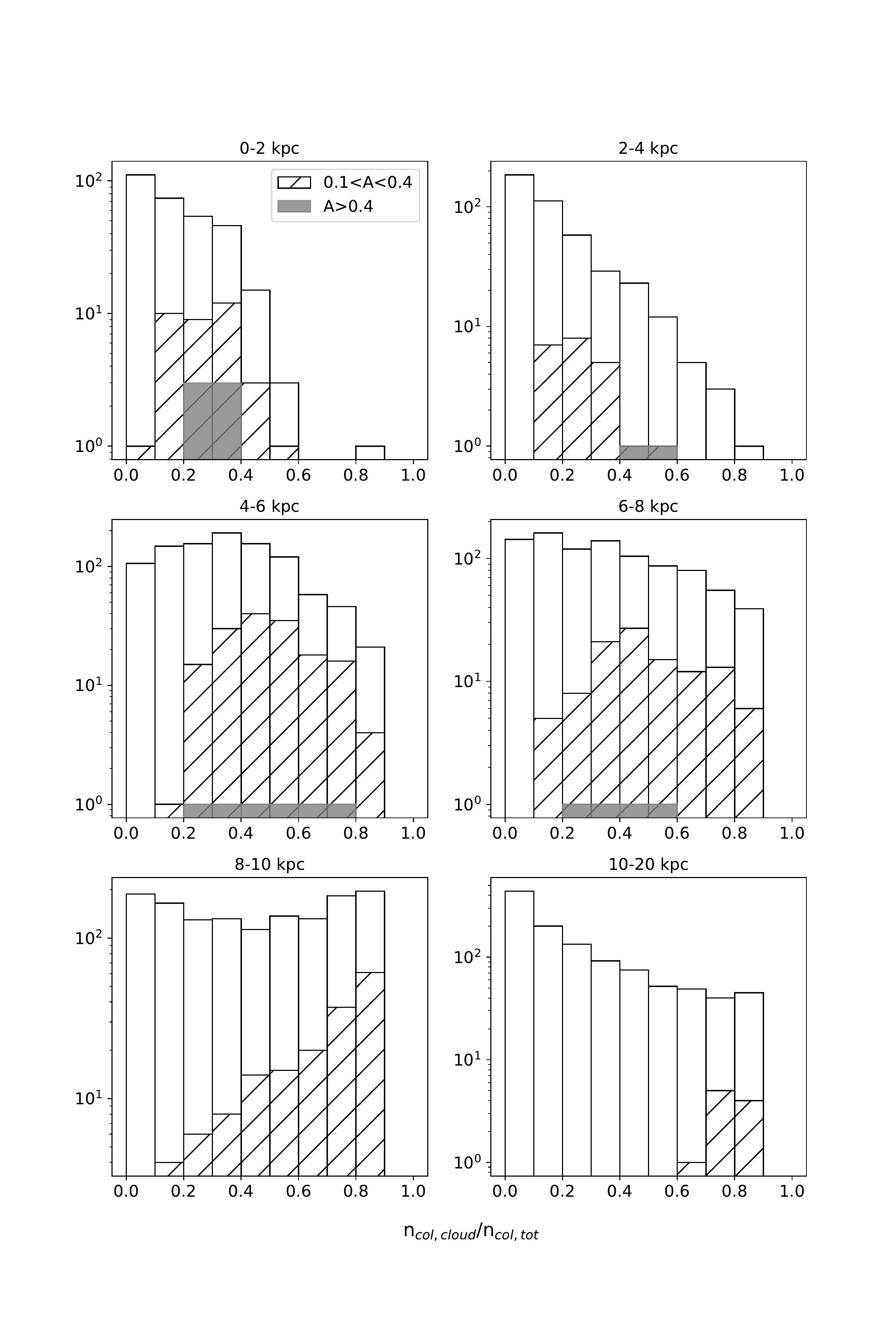}
\caption{Distribution of the ratio of the cloud's column density to the total column density in the direction of the cloud for different intervals of
distances from the Galactic Center. Only the clouds not overlapping with Fermi-LAT $\gamma$-ray sources are shown. The clouds characterized by the parameter $A>0.4$ (solid grey) and  $0.1<A<0.4$ (hatched) are highlighted. }
\label{fig:histfree}
\end{figure}

We calculated the fraction of gas belonging to each cloud of the MD16 catalog relative to the total gas in the l.o.s. included in the area of the cloud from the brightness of the CO, $W_{CO}$:
\begin{equation}
\rho=\frac{\int_{l-\theta}^{l+\theta} dl \int_{b-\theta/2}^{b+\theta} db \int_{v-2\sqrt{2\log 2} \sigma_v}^{v+2\sqrt{2\log 2}\sigma_v}  dv ~ W_{CO}(l,b,c)}{\int_{l-\theta/2}^{l+\theta} dl \int_{b-\theta}^{b+\theta} db \int_{-\infty}^{+\infty}  dv ~ W_{CO}(l,b,c)} \, ,
\end{equation}
where $\theta$ is the angular size of the cloud, $v$ is the radial velocity, and $\sigma_v$ is the dispersion of the velocity profile assumed to be Gaussian. The results are shown in \ref{fig:histfree} for clouds not overlapping (within 1.1 $\theta$) with any cataloged Fermi-LAT source. It appears that only a handful of clouds satisfy the condition of detectability. The limiting factor is given by the threshold of $A>0.4$ imposed by the sensitivity of Fermi-LAT. More than 50 clouds with $0.2<A<0.4$ are the dominant objects on the line of sight, having $\rho>0.65$. This number rises to 200 if we lower the threshold to $A=0.1$.

\section{The potential of current gamma-ray instruments}
Fermi-LAT is a powerful large field-of-view gamma-ray detector with the best performance at GeV energies. It is well designed to explore extended galactic sources, particularly SNRs and PWNe. This also concerns GMCs; however, the sensitivity of Fermi-LAT is at the margin of detection of gamma-rays from only a handful GMCs unless the CR density in the vicinity of the clouds does not substantially exceed  the local level. Another problem is the energy coverage. Because of the steep spectrum and the limited detection area of Fermi-LAT, the detection of gamma-rays even from the most favorable "passive"  GMCs with $A \sim 1$ cannot be extended beyond 0.1~TeV. The range of TeV energies is the domain of ground-based detectors - Imaging Atmospheric Cherenkov Telescopes (IACTs) and air shower particle arrays. However, for the current detectors, particularly HESS and HAWC, GMCs illuminated by $J_\odot$ are not accessible. 

\begin{figure*}[!h]
    \centering
    \includegraphics[width=0.8 \linewidth]{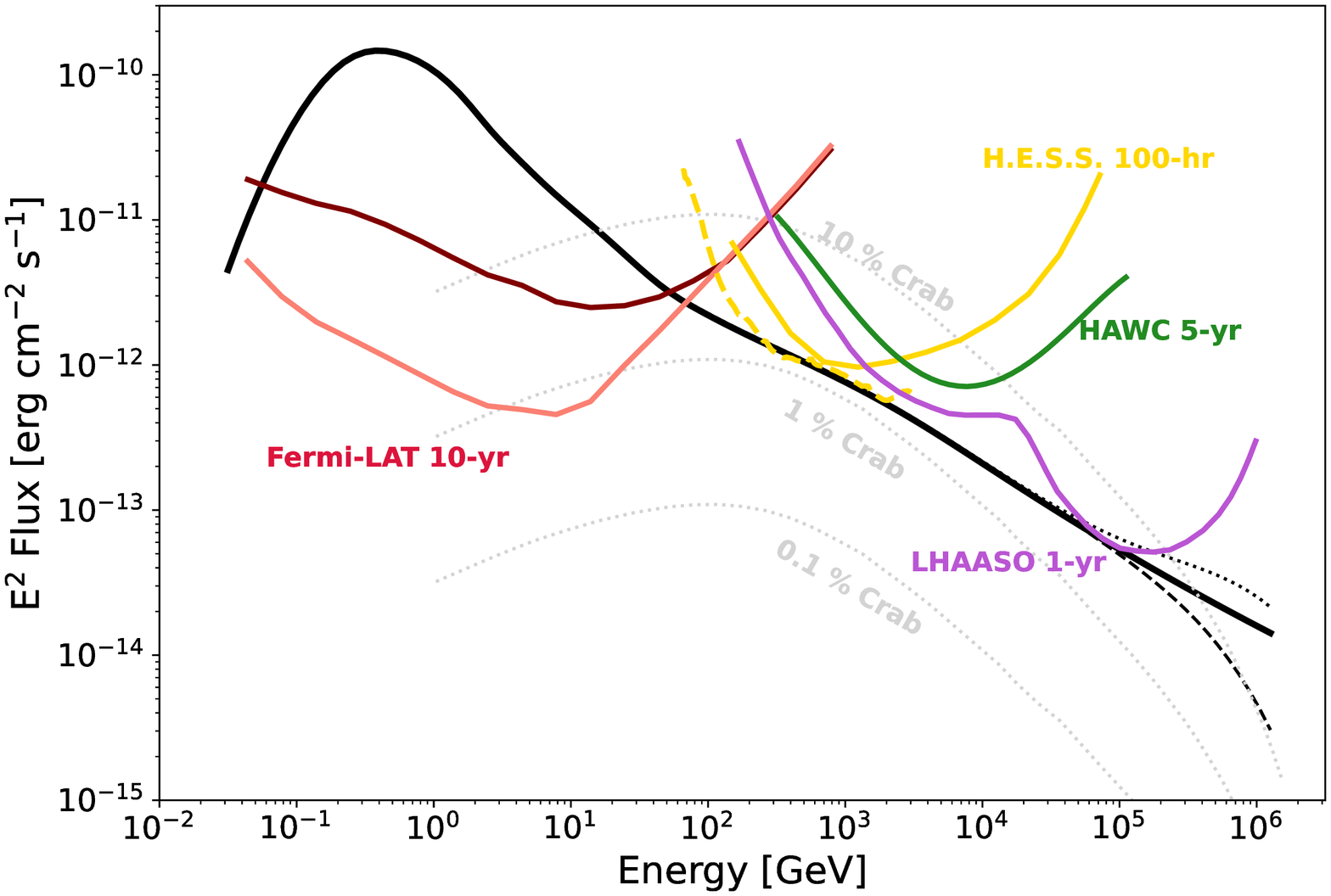}
    \caption{Gamma-ray fluxes expected from a GMC with A=1 compared to the point-like source sensitivities of currently operating $\gamma$-ray instruments: the 10-year Fermi-LAT sensitivity for the outer (light red) and inner Galaxy (dark red); the H.E.S.S. 5-telescope system sensitivity for 100 hr (solid yellow);  the HAWC 5-years sensitivity (green); and the LHAASO 1-year sensitivity (solid violet). The black solid line is the flux of a molecular cloud of $A=1$ illuminated by the "CR Sea". The two spectra above 10$^4$ GeV represents two options of CR proton spectra
    based on the spectra reported above $10^6$~GeV by the KASCADE (dashed line), and ICEtop (dotted line) collaborations (see the text).  As a reference, the fluxes representing 10, 1 and 0.1 percent of the gamma-ray flux from the Crab Nebula \citep{Cao2021Peta-electronNebula}  are shown.}
    \label{fig:current-sens}
\end{figure*}

This can be seen in Fig. \ref{fig:current-sens} where the flux sensitivities of the currently operating detectors are shown together with the flux induced by the \ {local CR Sea on a cloud with $A=1$, calculated with Eq.(\ref{eq:fluxmc})}.  In the plot are displayed: the sensitivity achieved after 10 years observations with Fermi-LAT for the inner ($l,b=(0,0)$; dark red) and outer ($l,b=(0,30)$; light red) Galaxy \citep{Maldera2019};  the H.E.S.S. sensitivity for 100 hours observation with the 4-telescopes configuration (solid yellow) \citep{Funk2013} and the preliminary calculation of the sensitivity for the 5-telescopes configuration (yellow dashed; \cite{Holler2015}); the HAWC sensitivity for 5 years of observations (green;\cite{Observatory}); and the LHAASO sensitivity for 1-year observations (magenta; \cite{DiSciascio2016}). The $\gamma$-ray flux of the given cloud exceeds the sensitivity of current instruments only at GeV energies. 

The condition for visibility of a molecular cloud can be determined by imposing that the flux of a cloud is higher or equal than the sensitivity, $S(t,E)$, calculated for a certain exposure time, $t$:

\begin{align}
F(E) & \geq S(t,E) \\
A \phi_\gamma (E) & \geq \sqrt{\frac{t_0}{t}}S_0(E)\\
A \sqrt{\frac{t}{t_0}} &\geq  \frac{S_0(E)}{\phi_{\gamma}(E)} \equiv \mathcal{R}_0(E)
\end{align}
here $S_0$ is the sensitivity calculated at a specific exposure $t_0$. In this sense $\mathcal{R}_0$ represents a condition for visibility as it is the minimum ratio to detect a cloud of $A=1$, which is characterized by a emissivity $\phi_{\gamma}$ with an instrument of sensitivity $S_0(E)$ calculated for a $t_0$-long exposure. 
  
The values for $\mathcal{R}_0$ for the current $\gamma$-ray instruments are plotted (dotted curves) in Fig \ref{fig:r-value}, \ {for the assumption of the local gamma-ray emissivity $\phi_\gamma=\phi_\gamma(J_\odot)$}. One can see that in order to measure \ {a similar emissivity as the local one} with the current TeV instruments, at least a $\mathcal{R}_0$ of $\sim$ 3 should be obtained. No single cloud in the Galaxy is characterized by $A \sim 3$,  except for some of the Gould Belt's clouds. However,  these nearby clouds are very extended, thus the sensitivity is significantly reduced. The sensitivity for a source of extension $\theta$ compared to the point-like source is worsened by the factor:
\begin{equation}
\omega(E,\theta) = \frac{\sqrt{\theta^2+\sigma^2_{PSF}(E)}}{\sigma_{PSF}(E)} \, ,
\end{equation}
where $\sigma_{PSF}$ is the instrument's point spread function. This results in a stricter condition on the visibility factor:
\begin{equation}
\mathcal{R}(E,\theta) = \omega(\theta,E) \mathcal{R}_0 = \omega(\theta,E) \frac{S_0(E)}{F_0(E)}	
\end{equation}
The worsening is especially significant for imaging air Cherenkov telescopes (IACTs) having the best angular resolution of 0.05-0.1$^\circ$ or better (see the top panel of Fig \ref{fig:sens-fact}). The effect is less dramatic for water Cherenkov (WC) detectors, which have a point spread function (PSF) of 0.1-0.3$^{\circ}$, comparable to the Fermi-LAT one and to the typical angular extensions of most of the clouds in the Galactic plane. 
  
The exposure time is another important factor for the detectability of GMCs. IACTs are  pointed telescopes with a small (a few degrees) FoV, while WC detectors cover simultaneously a significant fraction of the sky. The typical exposure time for specific segments of the Galactic Plane during the survey of the latter by IACTs over several years could be as large as 100 hours, which however is not sufficient to detect TeV gamma-rays from "passive" GMCs if illuminated by the local CR flux.  

The exposure time of a large fraction of clouds in the Galaxy by large FoV ground-based air shower particle detectors is much larger; it can be as large as 2000 hours per year (approximately 6 hours per night). Nevertheless, to observe passive clouds with HAWC, at least a factor $\mathcal{R}=10$ is needed, which is too large to be reached only with the exposure time increase.

\begin{figure*}[!h]
    \centering
    \includegraphics[width=0.8 \linewidth]{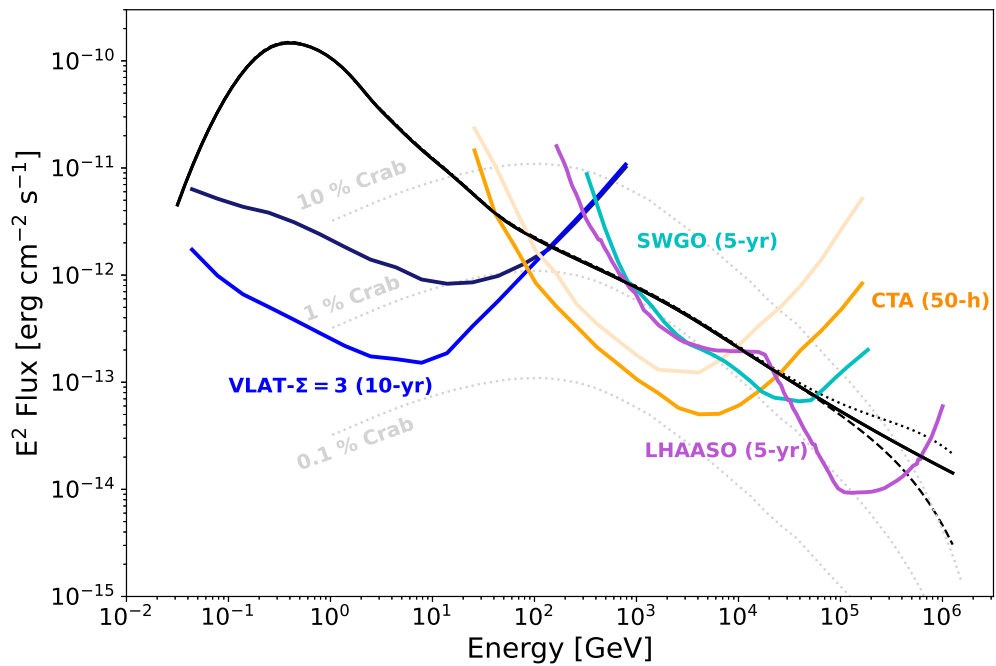}
    \caption{Point source sensitivity which will be obtained with future $\gamma$-ray instruments. We included: the sensitivity for an hypothetical Fermi Very Large Area Telescope (VLAT) with a 3 times larger effective area (blue curves). The expected sensitivity of CTA from the Northern (light orange) and Southern (orange) site, for 50 hours observations; the sensitivity of SWGO (cyan) after 5 years of osbservations and the sensitivity of LHAASO (magenta) for 5-years of observations.  }
\label{fig:future-sens}
\end{figure*}

\begin{figure}
 \includegraphics[width= 1\linewidth]{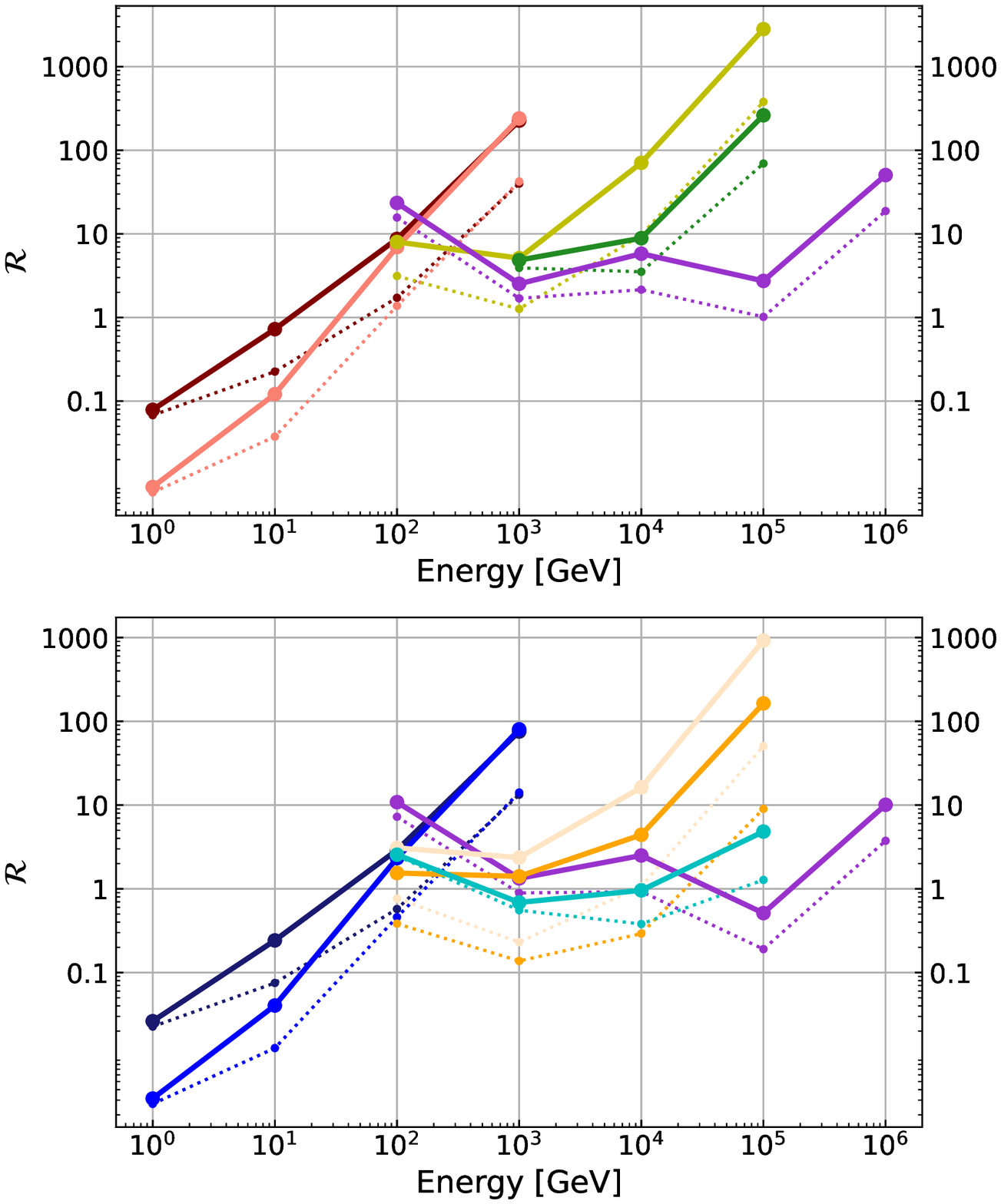}
 \caption{$\mathcal{R}$ values calculated for the current instruments (upper panel): Fermi-LAT (red), H.E.S.S. (yellow),HAWC (green), LHAASO (1-year; violet) and next-generation instruments (lower panel): VLAT (blue), CTA (orange), SWGO (cyan), LHAASO (5-years; violet). The exposure times are the same as in Figs \ref{fig:current-sens} and \ref{fig:future-sens}. The curves refer to the point source hypothesis (dotted lines) and to a 0.5$^\circ$-wide source  (solid lines).}
 \label{fig:r-value}
 \end{figure}

The recently completed  Water Cherenkov Detector Array (WCDA) of LHAASO will be able to detect  a limited number of passive GMCs  between 1 and 10 TeV. A breakthrough is expected at higher energies, thanks to the superior sensitivity of the KM2A (square km array) of LHAASO. For KM2A, $\mathcal{R} \lesssim 1$ after five years of observations, making the ultrahigh energies as effective  as  Fermi-LAT at GeV energies,  for searches of gamma-ray emitting clouds  (see Fig.\ref{fig:future-sens}).

\subsection{Analysis technique}
The recent advent of the \texttt{gammapy} software package \cite{Deil2017Gammapy:Astronomy} allows us to apply the 3D-likelihood technique,  initially used in the analysis of GeV data, also to the study of very-high-energy data. This is an essential tool for faint and extended gamma-ray sources.  Compared to the traditional methods \citep{Berge2007Background-rayastronomy} based on the subtraction of the background calculated in an ‘off’ region, the 3D likelihood analysis has certain advantages. It allows separation of the background caused by the hadronic showers from the large scale diffuse emission and, therefore, is more sensitive to low-surface-brightness sources with fluxes comparable to the diffuse gamma-ray emission of the galactic disk. Besides, the simultaneous fitting of all sources in the region of interest and the possibility of defining a spatial template based on observations on other wavelengths helps to resolve the cases of crowded areas.  Note that the exploitation of these factors leaded to the \ {first successful detection of a passive GMC at} TeV energies \citep{Sinha2021SearchSinha}.
 
\subsection{Runaway cosmic rays}
The detectability of a cloud in $\gamma$-rays  is significantly  improved in the case of location of the cloud in the environment with enhanced CR density caused by the presence in the proximity of recent and currently operating accelerator(s). 
Runaway CRs, i.e. particles which already have left the accelerator and injected into the circumstellar medium have been registered  both at GeV and TeV energies in the vicinity of some  middle-aged supernova remnants, (e.g. W44 \citep{Peron2020}, W28 \citep{Aharonian2008DiscoveryField}). The spectrum of runaway particles close/inside the clouds is hard to predict as it depends on different conditions such as the age of the accelerator, the distance of the clouds and of the diffusion coefficient \cite{Aharonian1996}. Meanwhile, the flux can be enhanced, e.g.  in the surroundings of W44 \citep{Peron2020}, by order of magnitude compared to the local CR flux.  If the injection occurs in a continuous regime, as in the case of massive star clusters, the CR density is expected to be strongly peaked towards the  the accelerator, therefore could be enhanced around the latter by orders of magnitude \citep{Aharonian2019}. Observations of the escaped particles are fundamental to understand the entire acceleration power of a source \citep{Gabici2007} \ {and new and future $\gamma$-ray instruments will help in constraining the spectrum of escaped particles at the highest energies.}

\section{The prospects}
The next generation of instruments will include the Cherenkov telescope array (CTA) and the Southern wide-field gamma-ray observatory (SWGO). The first will reach a sensitivity 10 times better than H.E.S.S., with an angular resolution close to 3 arcmins. Such an improved sensitivity would be promising to detect passive molecular clouds because in the case of CTA $\mathcal{R} \sim 0.5 $ at 1 TeV. The improved sensitivity, together with the better angular resolution, make CTA an ideal instrument to study not only the spectral energy distribution but also to investigate the spatial distribution of CRs inside the cloud itself. SWGO as well will reach a sensitivity an order of magnitude better than HAWC therefore it could be a valid counterpart of WCDA-LHAASO in the southern hemisphere, where the most massive clouds are located.  \ {Nevertheless, even with the improved sensitivity of the forthcoming gamma-ray telescopes, the measurements will remain limited to a handful of clouds in the VHE regime.}

\begin{figure}
\centering
\includegraphics[width= 0.8\linewidth]{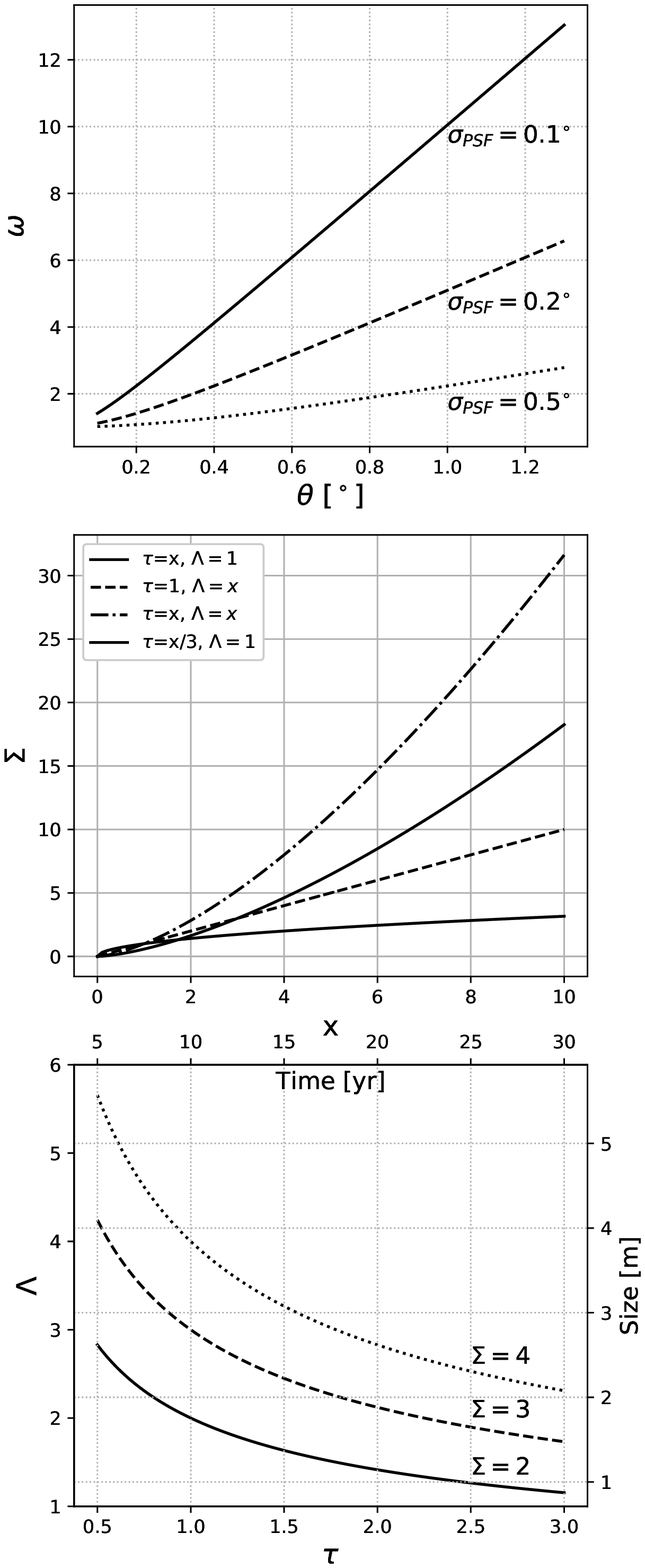}
\caption{Uppermost panel: the worsening factor of the sensitivity due to the source extension as a function of the extension for different angular resolutions. The middle and the lowermost panel show the combined effect of an extension, $\Lambda$, of the size and of an increase $\tau$ of the observation time on the total improvement, $\Sigma$, of the sensitivity.}
\label{fig:sens-fact}
\end{figure}

Meanwhile, even a relatively moderate (by a factor of 2-3) improvement of the  Fermi-LAT sensitivity at GeV energies would dramatically increase the number of clouds and thus provide a probe of the CR pressure (energy density) throughout the substantial fraction of the Galactic Disk.  Achieving such an improvement in sensitivity is not a trivial task. Given that GeV $\gamma$-rays are detected in the background-dominated regime, the minimum detectable flux (sensitivity) decreases with the exposure time as $t^{-1/2}$. Therefore, the resource of 12-year old Fermi-LAT in this regard is rather limited. Even assuming that Fermi-LAT will continue observations for another decade, the gain in the sensitivity cannot exceed 40 \%.  Clearly, one needs a new, more sensitive detector of GeV $\gamma$-rays. The improvement of sensitivity for the specific task of detection of $\gamma$-rays from extended GMCs cannot be realized by improving the angular resolution.  \ {Taking Fermi-LAT 10-yr sensitivity as reference, Fig. \ref{fig:sens-fact} shows how enlarging the exposure time by a factor $\tau$ or the size of the detector by a factor $\Lambda$ affects the sensitivity.} A breakthrough can be achieved only through an order of magnitude ($\Sigma > 3$) increase of the detection area. \ {Thus, we will need a new $Very$ large area telescope (VLAT) to achieve our goals. } This is a challenging but still feasible task for the space-based instruments; see, e.g. the recent proposal for the  Advanced Particle-astrophysics Telescope (APT) \citep{Buckley2019TheAPT}.
An improvement of the sensitivity of a factor $\Sigma \sim$3 can already be achieved by observing for 15 years with a Fermi-like instrument of size (3$\times$3) m$^2$. The APT  aims at having a $\sim$10 times larger effective area compared to the Fermi-LAT.  Two designs have been proposed: a (3$\times$3) m$^2$ instrument and a (3$\times$6) m$^2$ one. It is clear that, if this project will be approved, it will be an ideal instrument for our scopes.

\subsection{Probing the CR sea}
\begin{figure*}
\centering
\includegraphics[width=0.5 \linewidth]{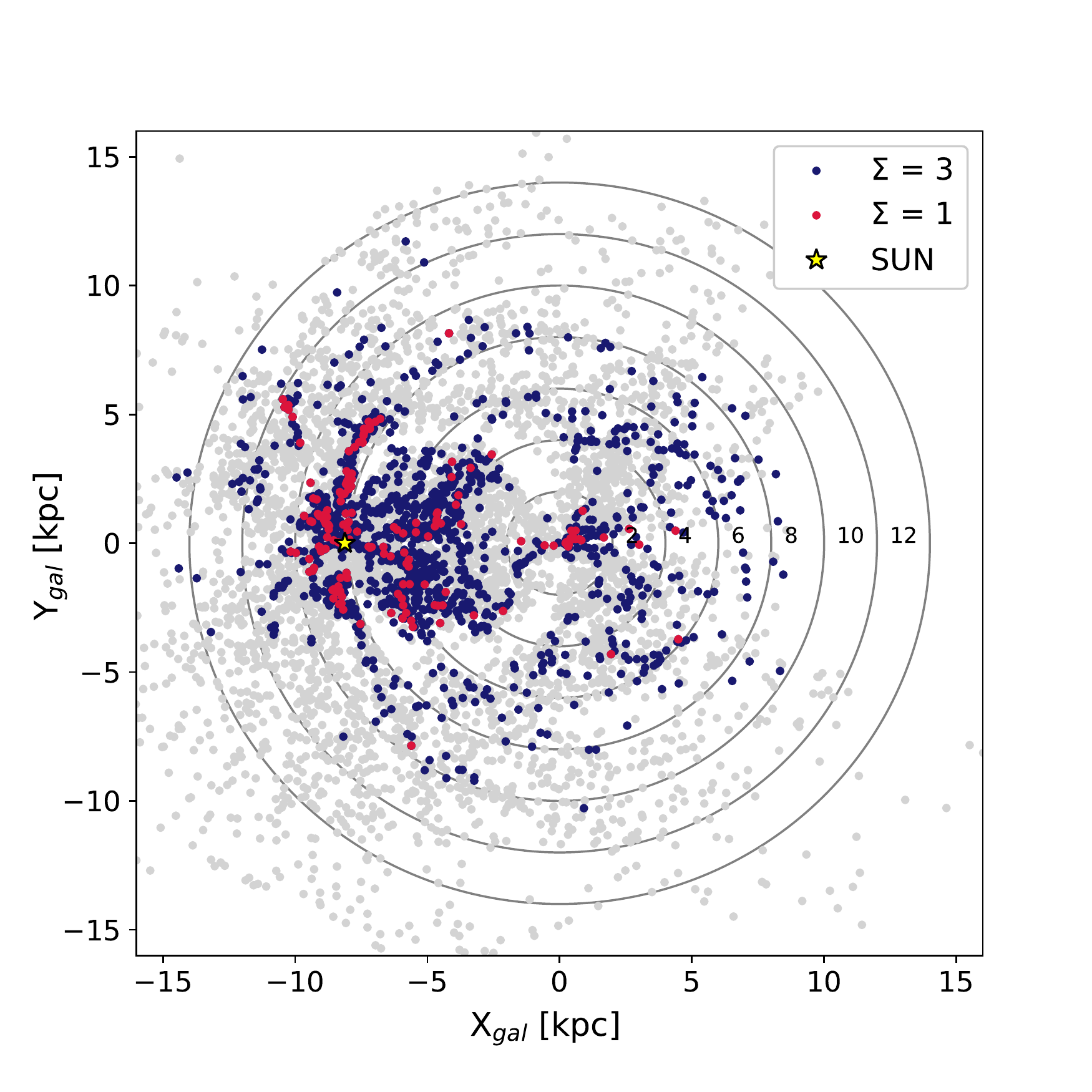}\includegraphics[width=0.5 \linewidth]{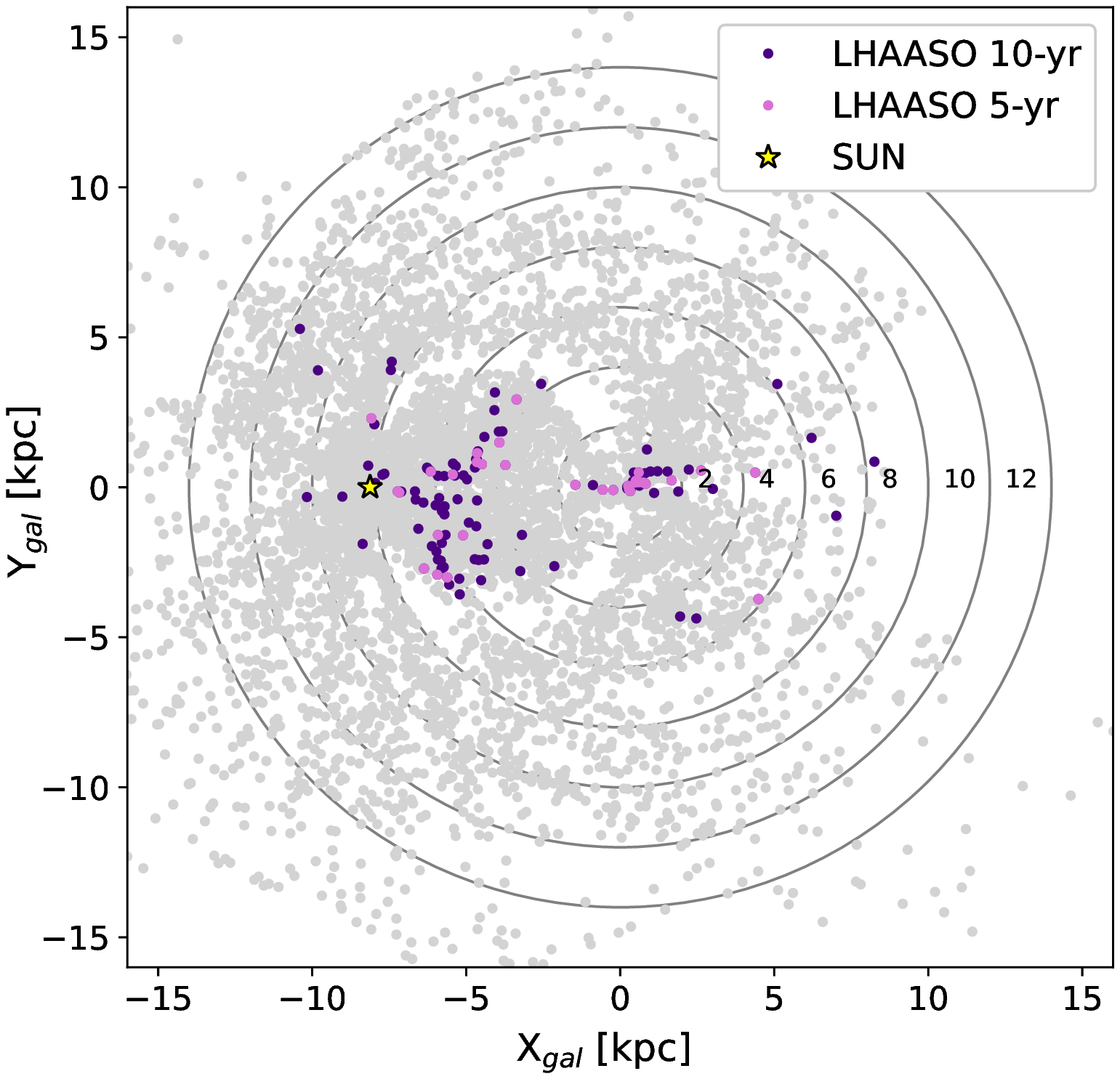}
\includegraphics[width=0.5 \linewidth]{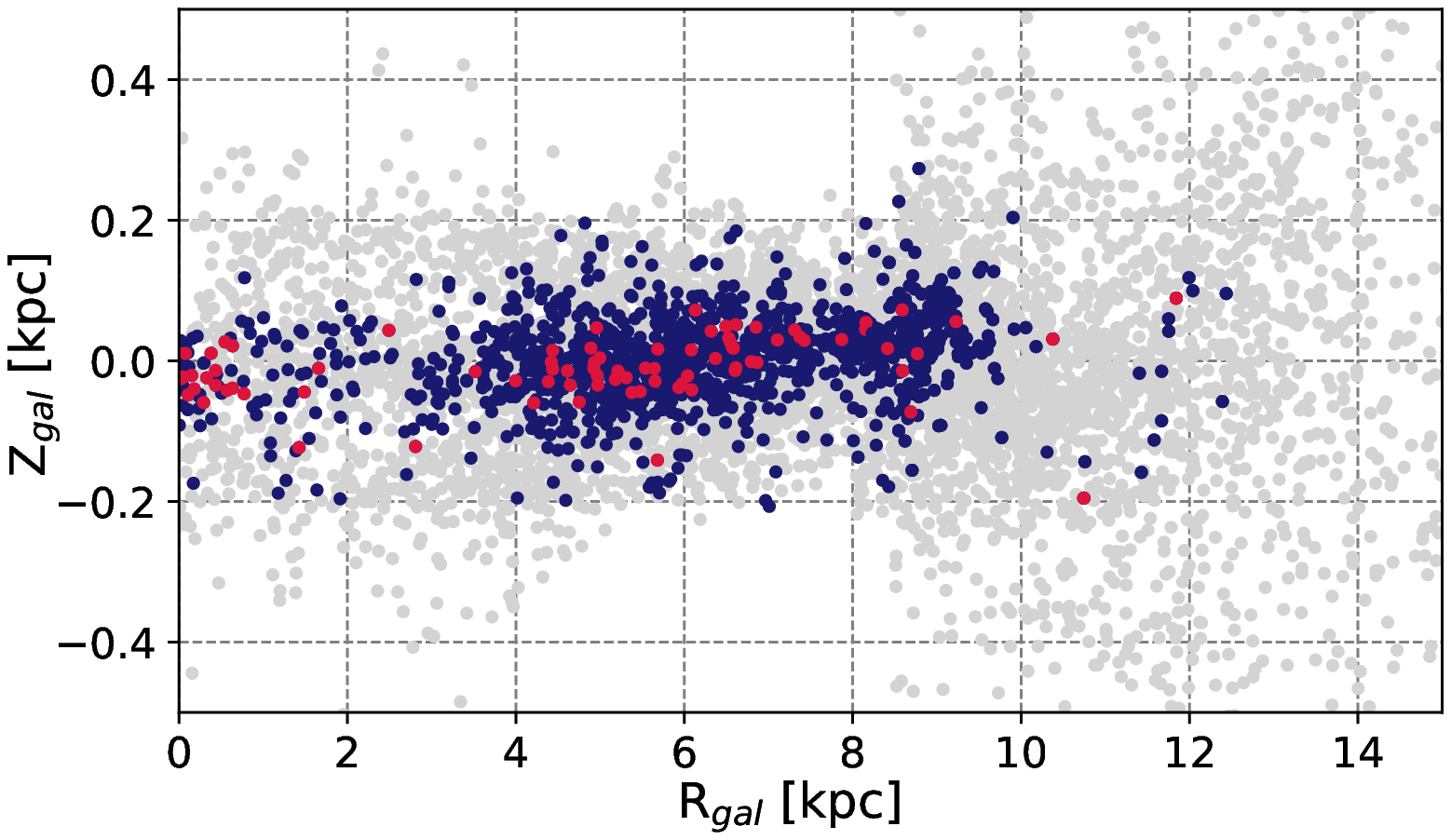}\includegraphics[width=0.5 \linewidth]{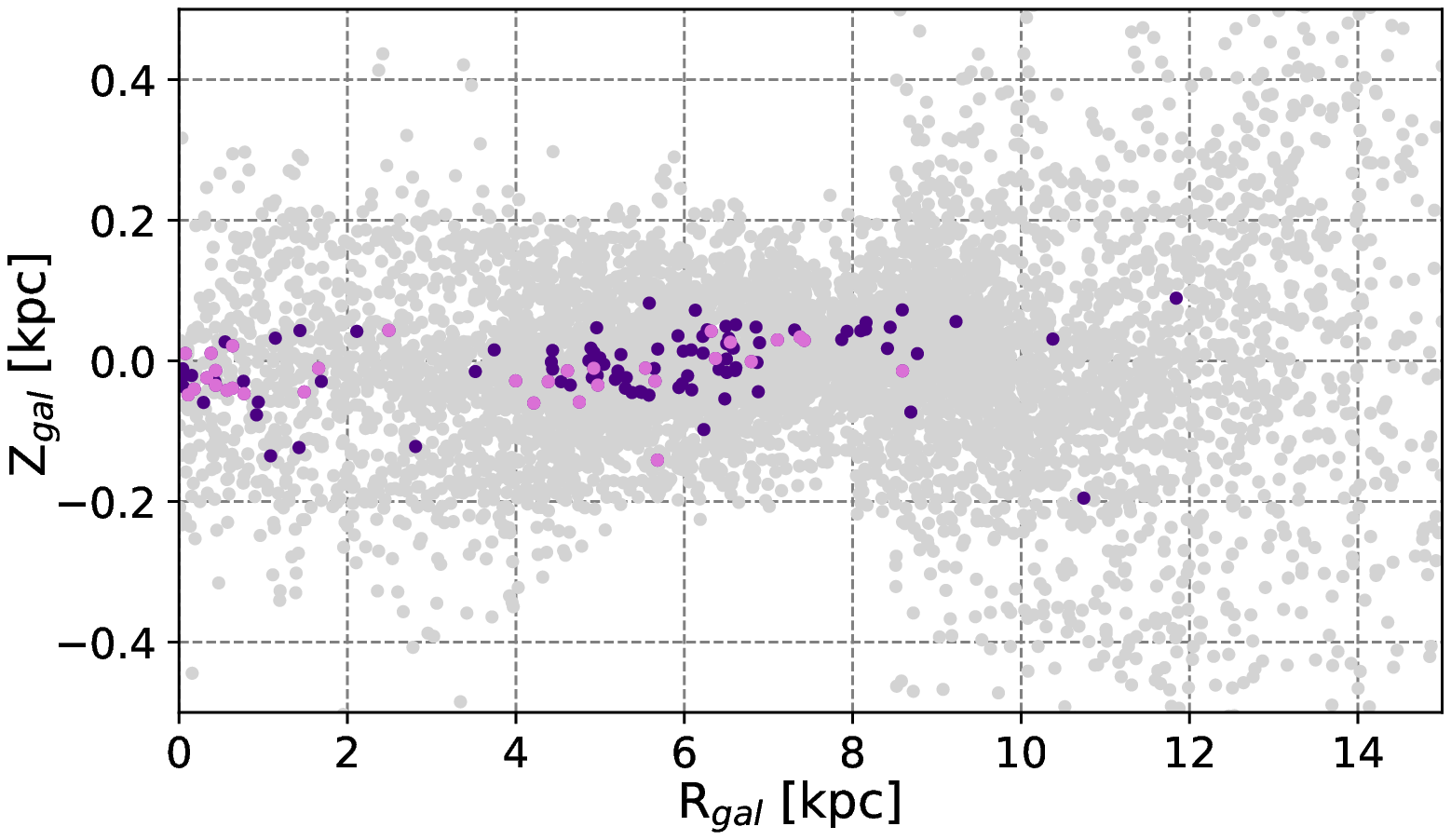}

\caption{Spatial distribution of the clouds from the MD16-catalog in the (X$_{gal}$,Y$_{gal}$) plane (upper panel) and in the ($R_{gal}$, $Z_{gal}$) plane (lower panel). In the left, the molecular clouds that overcome the detection threshold of Fermi-LAT and of an advanced detector with improved sensitivity $\Sigma=3$ are indicated in dark-red,  and blue, respectively. In the right, the clouds visible by LHAASO after 5 (light purple) and 10 (dark purple) years of observations are indicated. In the latter we assumed the same performance in the entire Galaxy, even though the sensitivity in the inner Galaxy should be worse. The size of the clouds is considered and an average angular resolution of 0.5$^\circ$ and 0.3$^\circ$ are assumed for Fermi-LAT and LHAASO, respectively.}
\label{fig:MC-map}
\end{figure*}

With its current sensitivity, the Fermi-LAT can map at most 1\% of the molecular clouds identified in the MD16-catalog. This corresponds to 40 objects, of which only 10 belong to the inner Galaxy. Lowering the sensitivity of a factor $\Sigma=3$ would increase the detectable sources to more than 1300 in total, of which $\sim$200 in the inner Galaxy ($<$ 4 kpc). The spatial distribution of the detectable MCs from the MD16-catalog is plotted in Fig. \ref{fig:MC-map}. With a factor  $\Sigma=3$ improvement, all galactic rings will be sampled with sufficient clouds, especially the 2--4 kpc ring, which is the most difficult to analyze because it is projected in a small range of longitudes, and therefore several sources may overlap.

Finally, while Fermi-LAT is limited to the observation of molecular clouds relatively close to the galactic plane, an advanced detector with improved sensitivity would allow access to several locations up to 400 parsecs above the plane (see the lower panel of Fig. \ref{fig:MC-map}). The combined knowledge of the cosmic-ray density at different distances from the Galactic Centre and at different heights from the Galactic plane would improve the knowledge regarding the propagation properties of these high-energy particles in the radial and perpendicular directions.

\section{Conclusions}

Gamma-ray emitting GMCs play a unique role of CR barometers allowing deep probes of the pressure (energy density) of CRs throughout the Galactic Disk with far-going astrophysical implications.  

The $\gamma$-ray fluxes from GMCs are faint and extended, which makes their detection difficult.  Yet, the analysis of Fermi-LAT observations of the Galactic Disk revealed $\gamma$-rays from a limited number of GMC in the energy interval between 1-100 GeV \citep{Aharonian2020, Peron2021,Baghmanyan2020}.   These results convincingly demonstrate the power of the method. At the same time,  they indicate that the potential of Fermi-LAT concerning the studies of GMCs is almost saturated.  For deeply probing CRs in different, including remote parts of the Galactic Disk, we need a new advanced $\gamma$-ray  detector in the GeV band (a "V-LAT") with sensitivity improved, compared to Fermi-LAT, by a factor of few. Hopefully, such a detector will appear in the foreseeable future.  \ {Such an instrument will be beneficial not only for effectively probing the CR "Sea" but also for searching of dark matter, investigating the nature of gamma-ray bursts and resolving other faint sources. }

Although with the increase of energy, the detection of  GMCs in $\gamma$-rays becomes more challenging, the CTA, as well as the water Cherenkov detectors like the proposed SWGO and currently operating WCDA-LHAASO,  should be able to detect  $\gamma$-rays in the energy interval between 1 to 10 TeV from GMCs characterised by the parameter $A \gtrsim 0.5$.   The domain of ultra-high energy $\gamma$-rays from 30 TeV up to 1 PeV looks even more promising.  One may predict that the recently completed and presently working in its full power KM2A-LHAASO in the coming years will detect GMCs in ultra-high energies and thus contribute significantly to uncovering the origin of highest energy CRs around the knee and beyond.

\bibliographystyle{aa} 
\bibliography{references} 

\begin{thebibliography}{44}
\expandafter\ifx\csname natexlab\endcsname\relax\def\natexlab#1{#1}\fi

\bibitem[{Aartsen {et~al.}(2019)Aartsen, Ackermann, Adams, Aguilar, Ahlers,
  Ahrens, Alispach, Andeen, Anderson, Ansseau, Anton, Arg{\"{u}}elles,
  Auffenberg, Axani, Backes, Bagherpour, Bai, Barbano, Barwick, Baum, Baur,
  Bay, Beatty, Becker, Becker~Tjus, Benzvi, Berley, Bernardini, Besson, Binder,
  Bindig, Blaufuss, Blot, Bohm, B{\"{o}}rner, B{\"{o}}ser, Botner,
  B{\"{o}}ttcher, Bourbeau, Bourbeau, Bradascio, Braun, Bretz, Bron,
  Brostean-Kaiser, Burgman, Buscher, Busse, Carver, Chen, Cheung, Chirkin,
  Clark, Classen, Collin, Conrad, Coppin, Correa, Cowen, Cross, Dave,
  De~Andr{\'{e}}, De~Clercq, Delaunay, Dembinski, Deoskar, De~Ridder, Desiati,
  De~Vries, De~Wasseige, De~With, Deyoung, Diaz, D{\'{i}}az-V{\'{e}}lez,
  Dujmovic, Dunkman, Dvorak, Eberhardt, Ehrhardt, Eller, Evenson, Fahey,
  Fazely, Felde, Feusels, Filimonov, Finley, Franckowiak, Friedman, Fritz,
  Gaisser, Gallagher, Ganster, Garrappa, Gerhardt, Ghorbani, Glauch,
  Gl{\"{u}}senkamp, Goldschmidt, Gonzalez, Grant, Griffith, G{\"{u}}nder,
  G{\"{u}}nd{\"{u}}z, Haack, Hallgren, Halve, Halzen, Hanson, Hebecker,
  Heereman, Heix, Helbing, Hellauer, Henningsen, Hickford, Hignight, Hill,
  Hoffman, Hoffmann, Hoinka, Hokanson-Fasig, Hoshina, Huang, Huber, Hultqvist,
  H{\"{u}}nnefeld, Hussain, In, Iovine, Ishihara, Jacobi, Japaridze, Jeong,
  Jero, Jones, Jonske, Joppe, Kang, Kappes, Kappesser, Karg, Karl, Karle, Katz,
  Kauer, Kelley, Kheirandish, Kim, Kintscher, Kiryluk, Kittler, Klein, Koirala,
  Kolanoski, K{\"{o}}pke, Kopper, Kopper, Koskinen, Kowalski, Krings,
  Kr{\"{u}}ckl, Kulacz, Kunwar, Kurahashi, Kyriacou, Labare, Lanfranchi,
  Larson, Lauber, Lazar, Leonard, Leuermann, Liu, Lohfink, Lozano~Mariscal, Lu,
  Lucarelli, L{\"{u}}nemann, Luszczak, Madsen, Maggi, Mahn, Makino, Mallik,
  Mallot, Mancina, Mari{\c{s}}, Maruyama, Mase, Maunu, Meagher, Medici, Medina,
  Meier, Meighen-Berger, Menne, Merino, Meures, Miarecki, Micallef,
  Moment{\'{e}}, Montaruli, Moore, Morse, Moulai, Muth, Nagai, Nahnhauer,
  Nakarmi, Naumann, Neer, Niederhausen, Nowicki, Nygren, Obertacke~Pollmann,
  Olivas, O'Murchadha, O'Sullivan, Palczewski, Pandya, Pankova, Park, Peiffer,
  P{\'{e}}rez De Los~Heros, Philippen, Pieloth, Pinat, Pizzuto, Plum, Porcelli,
  Price, Przybylski, Raab, Raissi, Rameez, Rauch, Rawlins, Rea, Reimann,
  Relethford, Renzi, Resconi, Rhode, Richman, Robertson, Rongen, Rott, Ruhe,
  Ryckbosch, Rysewyk, Safa, Sanchez~Herrera, Sandrock, Sandroos, Santander,
  Sarkar, Sarkar, Satalecka, Schaufel, Schlunder, Schmidt, Schneider,
  Schneider, Schumacher, Sclafani, Seckel, Seunarine, Shefali, Silva, Snihur,
  Soedingrekso, Soldin, Song, Spiczak, Spiering, Stachurska, Stamatikos,
  Stanev, Stasik, Stein, Stettner, Steuer, Stezelberger, Stokstad,
  St{\"{o}}{\ss}l, Strotjohann, St{\"{u}}rwald, Stuttard, Sullivan, Sutherland,
  Taboada, Tenholt, Ter-Antonyan, Terliuk, Tilav, Tomankova, T{\"{o}}nnis,
  Toscano, Tosi, Tselengidou, Tung, Turcati, Turcotte, Turley, Ty, Unger,
  Unland~Elorrieta, Usner, Vandenbroucke, Van~Driessche, Van~Eijk,
  Van~Eijndhoven, Vanheule, Van~Santen, Vraeghe, Walck, Wallace, Wallraff,
  Wandkowsky, Watson, Weaver, Weiss, Weldert, Wendt, Werthebach, Westerhoff,
  Whelan, Whitehorn, Wiebe, Wiebusch, Wille, Williams, Wills, Wolf, Wood, Wood,
  Woschnagg, Wrede, Xu, Xu, Xu, Yanez, Yodh, Yoshida, Yuan, \&
  Z{\"{o}}cklein}]{Aartsen2019}
Aartsen, M.~G., Ackermann, M., Adams, J., {et~al.} 2019, Physical Review D,
  100, 46

\bibitem[{Abdalla {et~al.}(2018)Abdalla, Abramowski, Aharonian, Ait~Benkhali,
  Ang{\"{u}}ner, Arakawa, Arrieta, Aubert, Backes, Balzer, Barnard, Becherini,
  Becker~Tjus, Berge, Bernhard, Bernl{\"{o}}hr, Blackwell, B{\"{o}}ttcher,
  Boisson, Bolmont, Bonnefoy, Bordas, Bregeon, Brun, Brun, Bryan,
  B{\"{u}}chele, Bulik, Capasso, Carrigan, Caroff, Carosi, Casanova, Cerruti,
  Chakraborty, Chaves, Chen, Chevalier, Colafrancesco, Condon, Conrad, Davids,
  Decock, Deil, Devin, De~Wilt, Dirson, Djannati-Ata{\"{i}}, Domainko, Donath,
  Drury, Dutson, Dyks, Edwards, Egberts, Eger, Emery, Ernenwein, Eschbach,
  Farnier, Fegan, Fernandes, Fiasson, Fontaine, F{\"{o}}rster, Funk,
  F{\"{u}}{\ss}ling, Gabici, Gallant, Garrigoux, Gast, Gat{\'{e}}, Giavitto,
  Giebels, Glawion, Glicenstein, Gottschall, Grondin, Hahn, Haupt, Hawkes,
  Heinzelmann, Henri, Hermann, Hinton, Hofmann, Hoischen, Holch, Holler, Horns,
  Ivascenko, Iwasaki, Jacholkowska, Jamrozy, Jankowsky, Jankowsky, Jingo,
  Jouvin, Jung-Richardt, Kastendieck, Katarzy{\'{n}}ski, Katsuragawa, Katz,
  Kerszberg, Khangulyan, Kh{\'{e}}lifi, King, Klepser, Klochkov,
  Kl{\'{u}}zniak, Komin, Kosack, Krakau, Kraus, Kr{\"{u}}ger, Laffon, Lamanna,
  Lau, Lees, Lefaucheur, Lemi{\`{e}}re, Lemoine-Goumard, Lenain, Leser, Lohse,
  Lorentz, Liu, L{\'{o}}pez-Coto, Lypova, Marandon, Malyshev, Marcowith,
  Mariaud, Marx, Maurin, Maxted, Mayer, Meintjes, Meyer, Mitchell, Moderski,
  Mohamed, Mohrmann, Mor{\aa}, Moulin, Murach, Nakashima, De~Naurois,
  Ndiyavala, Niederwanger, Niemiec, Oakes, O'Brien, Odaka, Ohm, Ostrowski, Oya,
  Padovani, Panter, Parsons, Paz~Arribas, Pekeur, Pelletier, Perennes,
  Petrucci, Peyaud, Piel, Pita, Poireau, Poon, Prokhorov, Prokoph,
  P{\"{u}}hlhofer, Punch, Quirrenbach, Raab, Rauth, Reimer, Reimer, Renaud,
  De~Los~Reyes, Rieger, Rinchiuso, Romoli, Rowell, Rudak, Rulten, Safi-Harb,
  Sahakian, Saito, Sanchez, Santangelo, Sasaki, Schandri, Schlickeiser,
  Sch{\"{u}}ssler, Schulz, Schwanke, Schwemmer, Seglar-Arroyo, Settimo,
  Seyffert, Shafi, Shilon, Shiningayamwe, Simoni, Sol, Spanier, Spir-Jacob,
  {Stawarz}, Steenkamp, Stegmann, Steppa, Sushch, Takahashi, Tavernet,
  Tavernier, Taylor, Terrier, Tibaldo, Tiziani, Tluczykont, Trichard, Tsirou,
  Tsuji, Tuffs, Uchiyama, Van Der~Walt, Van~Eldik, Van~Rensburg, Van~Soelen,
  Vasileiadis, Veh, Venter, Viana, Vincent, Vink, Voisin, V{\"{o}}lk,
  Vuillaume, Wadiasingh, Wagner, Wagner, Wagner, White, Wierzcholska, Willmann,
  W{\"{o}}rnlein, Wouters, Yang, Zaborov, Zacharias, Zanin, Zdziarski, Zech,
  Zefi, Ziegler, Zorn, \& Zywucka}]{Abdalla2018}
Abdalla, H., Abramowski, A., Aharonian, F., {et~al.} 2018, Astronomy and
  Astrophysics, 612, A1

\bibitem[{Abeysekara {et~al.}(2021)Abeysekara, Albert, Alfaro, Alvarez,
  Camacho, Arteaga-Velazquez, Arunbabu, Rojas, Solares, Baghmanyan,
  Belmont-Moreno, BenZvi, Blandford, Brisbois, Caballero-Mora, Capistran,
  Carraminana, Casanova, Cotti, de~Leon, De~la Fuente, Hernandez, Dingus,
  DuVernois, Durocher, Diaz-Velez, Ellsworth, Engel, Espinoza, Fan, Fang,
  Fleischhack, Fraija, Galvan-Gamez, Garcia, Garcıa-Gonzalez, Garfias,
  Giacinti, Gonzalez, Goodman, Harding, Hernandez, Hinton, Hona, Huang,
  Hueyotl-Zahuantitla, Huntemeyer, Iriarte, Jardin-Blicq, Joshi, Kieda, Lara,
  Lee, Vargas, Linnemann, Longinotti, Luis-Raya, Lundeen, Malone, Martinez,
  Martinez-Castellanos, Martinez-Castro, Matthews, Miranda-Romagnoli, Soto,
  Moreno, Mostafa, Nayerhoda, Nellen, Newbold, Nisa, Noriega-Papaqui,
  Olivera-Nieto, Omodei, Peisker, Araujo, Perez-Perez, Ren, Rho, Rosa-Gonzalez,
  Ruiz-Velasco, Salazar, Greus, Sandova, Schneider, Schoorlemmer, Serna, Smith,
  Springer, Surajbali, Tollefson, Torres, Torres-Escobedo, Urena-Mena,
  Weisgarber, Werner, Willox, Zepeda, Zhou, on, \& Alvarez}]{Abeysekara2021}
Abeysekara, A.~U., Albert, A., Alfaro, R., {et~al.} 2021, Nature Astronomy, 1

\bibitem[{Acero {et~al.}(2016)Acero, Ackermann, Ajello, Albert, Baldini,
  Ballet, Barbiellini, Bastieri, Bellazzini, Bissaldi, Bloom, Bonino,
  Bottacini, Brandt, Bregeon, Bruel, Buehler, Buson, Caliandro, Cameron,
  Caragiulo, Caraveo, Casandjian, Cavazzuti, Cecchi, Charles, Chekhtman,
  Chiang, Chiaro, Ciprini, Claus, Cohen-Tanugi, Conrad, Cuoco, Cutini,
  D'Ammando, de~Angelis, de~Palma, Desiante, Digel, Di~Venere, Drell, Favuzzi,
  Fegan, Ferrara, Focke, Franckowiak, Funk, Fusco, Gargano, Gasparrini,
  Giglietto, Giordano, Giroletti, Glanzman, Godfrey, Grenier, Guiriec, Hadasch,
  Harding, Hayashi, Hays, Hewitt, Hill, Horan, Hou, Jogler, J{\'{o}}hannesson,
  Kamae, Kuss, Landriu, Larsson, Latronico, Li, Li, Longo, Loparco, Lovellette,
  Lubrano, Maldera, Malyshev, Manfreda, Martin, Mayer, Mazziotta, McEnery,
  Michelson, Mirabal, Mizuno, Monzani, Morselli, Nuss, Ohsugi, Omodei, Orienti,
  Orlando, Ormes, Paneque, Pesce-Rollins, Piron, Pivato, Rain{\`{o}}, Rando,
  Razzano, Razzaque, Reimer, Reimer, Remy, Renault, S{\'{a}}nchez-Conde,
  Schaal, Schulz, Sgr{\`{o}}, Siskind, Spada, Spandre, Spinelli, Strong, Suson,
  Tajima, Takahashi, Thayer, Thompson, Tibaldo, Tinivella, Torres, Tosti,
  Troja, Vianello, Werner, Wood, Wood, Zaharijas, \& Zimmer}]{Acero2016}
Acero, F., Ackermann, M., Ajello, M., {et~al.} 2016, The Astrophysical Journal
  Supplement Series, 223

\bibitem[{Ade {et~al.}(2011)Ade, Aghanim, Arnaud, \& Ashdown}]{Ade2011}
Ade, P., Aghanim, N., Arnaud, M., \& Ashdown, M. 2011, Astronomy
  {\textbackslash}{\&} Astrophysics, 536, 16

\bibitem[{Aguilar {et~al.}(2015)Aguilar, Aisa, Alpat, Alvino, Ambrosi, Andeen,
  Arruda, Attig, Azzarello, Bachlechner, Barao, Barrau, Barrin, Bartoloni,
  Basara, Battarbee, Battiston, Bazo, Becker, Behlmann, Beischer, Berdugo,
  Bertucci, Bigongiari, Bindi, Bizzaglia, Bizzarri, Boella, De~Boer, Bollweg,
  Bonnivard, Borgia, Borsini, Boschini, Bourquin, Burger, Cadoux, Cai, Capell,
  Caroff, Casaus, Cascioli, Castellini, Cernuda, Cerreta, Cervelli, Chae,
  Chang, Chen, Chen, Cheng, Chen, Cheng, Chou, Choumilov, Choutko, Chung,
  Clark, Clavero, Coignet, Consolandi, Contin, Corti, Gil, Coste, Creus,
  Crispoltoni, Cui, Dai, Delgado, Della~Torre, Demirk{\"{o}}z, Derome,
  Di~Falco, Di~Masso, Dimiccoli, D{\'{i}}az, Von~Doetinchem, Donnini, Du,
  Duranti, D'Urso, Eline, Eppling, Eronen, Fan, Farnesini, Feng, Fiandrini,
  Fiasson, Finch, Fisher, Galaktionov, Gallucci, Garc{\'{i}}a,
  Garc{\'{i}}a-L{\'{o}}pez, Gargiulo, Gast, Gebauer, Gervasi, Ghelfi, Gillard,
  Giovacchini, Goglov, Gong, Goy, Grabski, Grandi, Graziani, Guandalini,
  Guerri, Guo, Haas, Habiby, Haino, Han, He, Heil, Hoffman, Hsieh, Huang, Huh,
  Incagli, Ionica, Jang, Jinchi, Kanishev, Kim, Kim, Kirn, Kossakowski,
  Kounina, Kounine, Koutsenko, Krafczyk, La~Vacca, Laudi, Laurenti, Lazzizzera,
  Lebedev, Lee, Lee, Leluc, Levi, Li, Li, Li, Li, Li, Li, Li, Li, Li, Lim, Lin,
  Lipari, Lippert, Liu, Liu, Lolli, Lomtadze, Lu, Lu, Lu, Luebelsmeyer, Luo,
  Lv, Majka, Ma{\~{n}}{\'{a}}, Mar{\'{i}}n, Martin, Mart{\'{i}}nez, Masi,
  Maurin, Menchaca-Rocha, Meng, Mo, Morescalchi, Mott, M{\"{u}}ller, Ni,
  Nikonov, Nozzoli, Nunes, Obermeier, Oliva, Orcinha, Palmonari, Palomares,
  Paniccia, Papi, Pauluzzi, Pedreschi, Pensotti, Pereira, Picot-Clemente, Pilo,
  Piluso, Pizzolotto, Plyaskin, Pohl, Poireau, Postaci, Putze, Quadrani, Qi,
  Qin, Qu, R{\"{a}}ih{\"{a}}, Rancoita, Rapin, Ricol, Rodr{\'{i}}guez,
  Rosier-Lees, Rozhkov, Rozza, Sagdeev, Sandweiss, Saouter, Sbarra, Schael,
  Schmidt, Von~Dratzig, Schwering, Scolieri, Seo, Shan, Shan, Shi, Shi, Shi,
  Siedenburg, Son, Spada, Spinella, Sun, Sun, Tacconi, Tang, Tang, Tang, Tao,
  Tescaro, Ting, Ting, Tomassetti, Torsti, T{\"{u}}rkolu, Urban, Vagelli,
  Valente, Vannini, Valtonen, Vaurynovich, Vecchi, Velasco, Vialle, Vitale,
  Vitillo, Wang, Wang, Wang, Wang, Wang, Wang, Weng, Whitman,
  Wienkenh{\"{o}}ver, Wu, Wu, Xia, Xie, Xie, Xiong, Xin, Xu, Xu, Yan, Yang,
  Yang, Ye, Yi, Yu, Yu, Zeissler, Zhang, Zhang, Zhang, Zhang, Zheng, Zhuang,
  Zhukov, Zichichi, Zimmermann, Zuccon, \& Zurbach}]{Aguilar2015}
Aguilar, M., Aisa, D., Alpat, B., {et~al.} 2015, Physical Review Letters, 114

\bibitem[{Aharonian {et~al.}(2008)Aharonian, Akhperjanian, Bazer-Bachi, Behera,
  Beilicke, Benbow, Berge, Bernl{\"{o}}hr, Boisson, Bolz, Borrel, Braun, Brion,
  Brown, B{\"{u}}hler, Bulik, B{\"{u}}sching, Boutelier, Carrigan, Chadwick,
  Chounet, Clapson, Coignet, Cornils, Costamante, Degrange, Dickinson,
  Djannati-Ata{\"{i}}, Domainko, O'C.~Drury, Dubus, Dyks, Egberts,
  Emmanoulopoulos, Espigat, Farnier, Feinstein, Fiasson, F{\"{o}}rster,
  Fontaine, Fukui, Funk, Funk, F{\"{u}}{\ss}ling, Gallant, Giebels,
  Glicenstein, Gl{\"{u}}ck, Goret, Hadjichristidis, Hauser, Hauser,
  Heinzelmann, Henri, Hermann, Hinton, Hoffmann, Hofmann, Holleran, Hoppe,
  Horns, Jacholkowska, de~Jager, Kendziorra, Kerschhaggl, Kh{\'{e}}lifi, Komin,
  Kosack, Lamanna, Latham, Le~Gallou, Lemi{\`{e}}re, Lemoine-Goumard, Lenain,
  Lohse, Martin, Martineau-Huynh, Marcowith, Masterson, Maurin, McComb,
  Moderski, Moriguchi, Moulin, de~Naurois, Nedbal, Nolan, Olive, Orford,
  Osborne, Ostrowski, Panter, Pedaletti, Pelletier, Petrucci, Pita,
  P{\"{u}}hlhofer, Punch, Ranchon, Raubenheimer, Raue, Rayner, Reimer, Renaud,
  Ripken, Rob, Rolland, Rosier-Lees, Rowell, Rudak, Ruppel, Sahakian,
  Santangelo, Saug{\'{e}}, Schlenker, Schlickeiser, Schr{\"{o}}der, Schwanke,
  Schwarzburg, Schwemmer, Shalchi, Sol, Spangler, Stawarz, Steenkamp, Stegmann,
  Superina, Takeuchi, Tam, Tavernet, Terrier, van Eldik, Vasileiadis, Venter,
  Vialle, Vincent, Vivier, V{\"{o}}lk, Volpe, Wagner, Ward, Aharonian,
  Akhperjanian, Bazer-Bachi, Behera, Beilicke, Benbow, Berge, Bernl{\"{o}}hr,
  Boisson, Bolz, Borrel, Braun, Brion, Brown, B{\"{u}}hler, Bulik,
  B{\"{u}}sching, Boutelier, Carrigan, Chadwick, Chounet, Clapson, Coignet,
  Cornils, Costamante, Degrange, Dickinson, Djannati-Ata{\"{i}}, Domainko,
  O'C.~Drury, Dubus, Dyks, Egberts, Emmanoulopoulos, Espigat, Farnier,
  Feinstein, Fiasson, F{\"{o}}rster, Fontaine, Fukui, Funk, Funk,
  F{\"{u}}{\ss}ling, Gallant, Giebels, Glicenstein, Gl{\"{u}}ck, Goret,
  Hadjichristidis, Hauser, Hauser, Heinzelmann, Henri, Hermann, Hinton,
  Hoffmann, Hofmann, Holleran, Hoppe, Horns, Jacholkowska, de~Jager,
  Kendziorra, Kerschhaggl, Kh{\'{e}}lifi, Komin, Kosack, Lamanna, Latham,
  Le~Gallou, Lemi{\`{e}}re, Lemoine-Goumard, Lenain, Lohse, Martin,
  Martineau-Huynh, Marcowith, Masterson, Maurin, McComb, Moderski, Moriguchi,
  Moulin, de~Naurois, Nedbal, Nolan, Olive, Orford, Osborne, Ostrowski, Panter,
  Pedaletti, Pelletier, Petrucci, Pita, P{\"{u}}hlhofer, Punch, Ranchon,
  Raubenheimer, Raue, Rayner, Reimer, Renaud, Ripken, Rob, Rolland,
  Rosier-Lees, Rowell, Rudak, Ruppel, Sahakian, Santangelo, Saug{\'{e}},
  Schlenker, Schlickeiser, Schr{\"{o}}der, Schwanke, Schwarzburg, Schwemmer,
  Shalchi, Sol, Spangler, Stawarz, Steenkamp, Stegmann, Superina, Takeuchi,
  Tam, Tavernet, Terrier, van Eldik, Vasileiadis, Venter, Vialle, Vincent,
  Vivier, V{\"{o}}lk, Volpe, Wagner, \& Ward}]{Aharonian2008DiscoveryField}
Aharonian, F., Akhperjanian, A.~G., Bazer-Bachi, A.~R., {et~al.} 2008, A{\&}A,
  481, 401

\bibitem[{Aharonian {et~al.}(2020)Aharonian, Peron, Yang, Casanova, \&
  Zanin}]{Aharonian2020}
Aharonian, F., Peron, G., Yang, R., Casanova, S., \& Zanin, R. 2020, Physical
  Review D, 101

\bibitem[{Aharonian {et~al.}(2019)Aharonian, Yang, \&
  de~O{\~{n}}a~Wilhelmi}]{Aharonian2019}
Aharonian, F., Yang, R., \& de~O{\~{n}}a~Wilhelmi, E. 2019, Nature Astronomy,
  3, 561

\bibitem[{Aharonian \& Atoyan(1996)}]{Aharonian1996}
Aharonian, F.~A. \& Atoyan, A.~M. 1996, Astronomy {\&} Astrophysics, 309, 917

\bibitem[{Albert {et~al.}(2020)Albert, Alfaro, Alvarez, Camacho,
  Arteaga-Vel{\'{a}}zquez, Arunbabu, Avila~Rojas, Ayala~Solares, Baghmanyan,
  Belmont-Moreno, BenZvi, Brisbois, Caballero-Mora, Capistr{\'{a}}n,
  Carrami{\~{n}}ana, Casanova, Cotti, Couti{\~{n}}o~de Le{\'{o}}n, De~la
  Fuente, Diaz~Hernandez, Diaz-Cruz, Dingus, DuVernois, Durocher,
  D{\'{i}}az-V{\'{e}}lez, Ellsworth, Engel, Espinoza, Fan, Fang, Alonso,
  Fleischhack, Fraija, Galv{\'{a}}n-G{\'{a}}mez, Garcia,
  Garc{\'{i}}a-Gonz{\'{a}}lez, Garfias, Giacinti, Gonz{\'{a}}lez, Goodman,
  Harding, Hernandez, Hinton, Hona, Huang, Hueyotl-Zahuantitla,
  H{\"{u}}ntemeyer, Iriarte, Jardin-Blicq, Joshi, Kieda, Lara, Lee,
  Le{\'{o}}n~Vargas, Linnemann, Longinotti, Luis-Raya, Lundeen,
  L{\'{o}}pez-Coto, Malone, Marandon, Martinez, Martinez-Castellanos,
  Mart{\'{i}}nez-Castro, Matthews, Miranda-Romagnoli, Morales-Soto, Moreno,
  Mostaf{\'{a}}, Nayerhoda, Nellen, Newbold, Nisa, Noriega-Papaqui,
  Olivera-Nieto, Omodei, Peisker, P{\'{e}}rez~Araujo, P{\'{e}}rez-P{\'{e}}rez,
  Ren, Rho, Rivi{\`{e}}re, Rosa-Gonz{\'{a}}lez, Ruiz-Velasco, Salazar,
  Salesa~Greus, Sandoval, Schneider, Schoorlemmer, Serna, Sinnis, Smith,
  Springer, Surajbali, Tollefson, Torres, Torres-Escobedo, Ukwatta,
  Ure{\~{n}}a-Mena, Weisgarber, Werner, Willox, Zepeda, Zhou, de~Le{\'{o}}n,
  {\'{A}}lvarez, Collaboration, Albert, Alfaro, Alvarez, Camacho,
  Arteaga-Vel{\'{a}}zquez, Arunbabu, Avila~Rojas, Ayala~Solares, Baghmanyan,
  Belmont-Moreno, BenZvi, Brisbois, Caballero-Mora, Capistr{\'{a}}n,
  Carrami{\~{n}}ana, Casanova, Cotti, Couti{\~{n}}o~de Le{\'{o}}n, De~la
  Fuente, Diaz~Hernandez, Diaz-Cruz, Dingus, DuVernois, Durocher,
  D{\'{i}}az-V{\'{e}}lez, Ellsworth, Engel, Espinoza, Fan, Fang, Alonso,
  Fleischhack, Fraija, Galv{\'{a}}n-G{\'{a}}mez, Garcia,
  Garc{\'{i}}a-Gonz{\'{a}}lez, Garfias, Giacinti, Gonz{\'{a}}lez, Goodman,
  Harding, Hernandez, Hinton, Hona, Huang, Hueyotl-Zahuantitla,
  H{\"{u}}ntemeyer, Iriarte, Jardin-Blicq, Joshi, Kieda, Lara, Lee,
  Le{\'{o}}n~Vargas, Linnemann, Longinotti, Luis-Raya, Lundeen,
  L{\'{o}}pez-Coto, Malone, Marandon, Martinez, Martinez-Castellanos,
  Mart{\'{i}}nez-Castro, Matthews, Miranda-Romagnoli, Morales-Soto, Moreno,
  Mostaf{\'{a}}, Nayerhoda, Nellen, Newbold, Nisa, Noriega-Papaqui,
  Olivera-Nieto, Omodei, Peisker, P{\'{e}}rez~Araujo, P{\'{e}}rez-P{\'{e}}rez,
  Ren, Rho, Rivi{\`{e}}re, Rosa-Gonz{\'{a}}lez, Ruiz-Velasco, Salazar,
  Salesa~Greus, Sandoval, Schneider, Schoorlemmer, Serna, Sinnis, Smith,
  Springer, Surajbali, Tollefson, Torres, Torres-Escobedo, Ukwatta,
  Ure{\~{n}}a-Mena, Weisgarber, Werner, Willox, Zepeda, Zhou, de~Le{\'{o}}n,
  {\'{A}}lvarez, \& Collaboration}]{Albert20203HWC:Sources}
Albert, A., Alfaro, R., Alvarez, C., {et~al.} 2020, ApJ, 905, 76

\bibitem[{Amato \& Casanova(2021)}]{Amato2021}
Amato, E. \& Casanova, S. 2021, Journal of Plasma Physics, 87, 845870101

\bibitem[{Amenomori {et~al.}(2021)Amenomori, Bao, Bi, Chen, Chen, Chen, Chen,
  Chen, Cui, Ding, Fang, Fang, Feng, Feng, Feng, Gao, Gou, Guo, Guo, He, He,
  Hibino, Hotta, Hu, Hu, Huang, Jia, Jiang, Jin, Kasahara, Katayose, Kato,
  Kato, Kawata, Kihara, Ko, Kozai, Le, Li, Li, Li, Lin, Liu, Liu, Liu, Liu,
  Liu, Lou, Lu, Meng, Munakata, Nakada, Nakamura, Nanjo, Nishizawa, Ohnishi,
  Ohura, Ozawa, Qian, Qu, Saito, Sakata, Sako, Shao, Shibata, Shiomi, Sugimoto,
  Takano, Takita, Tan, Tateyama, Torii, Tsuchiya, Udo, Wang, Wu, Xue, Yamamoto,
  Yang, Yokoe, Yuan, Zhai, Zhang, Zhang, Zhang, Zhang, Zhang, Zhang, Zhang,
  Zhao, \& Zhou}]{Amenomori2021}
Amenomori, M., Bao, Y.~W., Bi, X.~J., {et~al.} 2021, Physical Review Letters,
  126, 141101

\bibitem[{Apel {et~al.}(2013)Apel, Arteaga-Vel{\'{a}}zquez, Bekk, Bertaina,
  Bl{\"{u}}mer, Bozdog, Brancus, Cantoni, Chiavassa, Cossavella, Daumiller,
  De~Souza, Di~Pierro, Doll, Engel, Engler, Finger, Fuchs, Fuhrmann, Gils,
  Glasstetter, Grupen, Haungs, Heck, H{\"{o}}randel, Huber, Huege, Kampert,
  Kang, Klages, Link, {\L}uczak, Ludwig, Mathes, Mayer, Melissas, Milke,
  Mitrica, Morello, Oehlschl{\"{a}}ger, Ostapchenko, Palmieri, Petcu, Pierog,
  Rebel, Roth, Schieler, Schoo, Schr{\"{o}}der, Sima, Toma, Trinchero, Ulrich,
  Weindl, Wochele, Wommer, \& Zabierowski}]{Apel2013}
Apel, W.~D., Arteaga-Vel{\'{a}}zquez, J.~C., Bekk, K., {et~al.} 2013,
  Astroparticle Physics, 47, 54

\bibitem[{Atri {et~al.}(2014)Atri, Melott, Atri, \&
  Melott}]{Atri2014CosmicReview}
Atri, D., Melott, A.~L., Atri, D., \& Melott, A.~L. 2014, APh, 53, 186

\bibitem[{Baghmanyan {et~al.}(2020)Baghmanyan, Peron, Casanova, Aharonian, \&
  Zanin}]{Baghmanyan2020}
Baghmanyan, V., Peron, G., Casanova, S., Aharonian, F., \& Zanin, R. 2020,
  Astrophysical Journal Letters, 901

\bibitem[{Berge {et~al.}(2007)Berge, Funk, \&
  Hinton}]{Berge2007Background-rayastronomy}
Berge, D., Funk, S., \& Hinton, J. 2007, Astronomy {\&} Astrophysics, 466, 1219

\bibitem[{Bolatto {et~al.}(2013)Bolatto, Wolfire, \& Leroy}]{Bolatto2013}
Bolatto, A.~D., Wolfire, M., \& Leroy, A.~K. 2013, Annual Review of Astronomy
  and Astrophysics, 51, 207

\bibitem[{Buckley {et~al.}(2019)Buckley, Bergstrom, Binns, Buhler, Chen,
  Cherry, Funk, Hooper, Mitchell, De~Nolfo, Nussirat, Profumo, Rauch, Stern,
  Varner, Wakely, Zink, Buckley, Bergstrom, Binns, Buhler, Chen, Cherry, Funk,
  Hooper, Mitchell, De~Nolfo, Nussirat, Profumo, Rauch, Stern, Varner, Wakely,
  \& Zink}]{Buckley2019TheAPT}
Buckley, J., Bergstrom, L., Binns, B., {et~al.} 2019, BAAS, 51, 78

\bibitem[{Cao(2021)}]{Cao2021AnM}
Cao, Z. 2021, Nature Astronomy 2021 5:8, 5, 849

\bibitem[{Cao {et~al.}(2021)Cao, Aharonian, Axikegu, Bai, Bai, Bastieri, \&
  Bi}]{Cao2021Peta-electronNebula}
Cao, Z., Aharonian, F., Axikegu, Q.~A., {et~al.} 2021, Science, 373, 425

\bibitem[{Dame {et~al.}(2000)Dame, Hartmann, \& Thaddeus}]{Dame2000}
Dame, T.~M., Hartmann, D., \& Thaddeus, P. 2000, The Astrophysical Journal,
  547, 792

\bibitem[{Deil {et~al.}(2017)Deil, Donath, Owen, Terrier, B{\"{u}}hler,
  Armstrong, Deil, Donath, Owen, Terrier, B{\"{u}}hler, \&
  Armstrong}]{Deil2017Gammapy:Astronomy}
Deil, C., Donath, A., Owen, E., {et~al.} 2017, Astrophysics Source Code
  Library, record ascl:1711.014, ascl:1711.014

\bibitem[{Di~Sciascio(2016)}]{DiSciascio2016}
Di~Sciascio, G. 2016, Nuclear and Particle Physics Proceedings, 279-281, 166

\bibitem[{Funk \& Hinton(2013)}]{Funk2013}
Funk, S. \& Hinton, J.~A. 2013, Astroparticle Physics, 43, 348

\bibitem[{Gabici {et~al.}(2007)Gabici, Aharonian, \& Blasi}]{Gabici2007}
Gabici, S., Aharonian, F.~A., \& Blasi, P. 2007, Astrophysics and Space
  Science, 309, 365

\bibitem[{{HAWC}(2020)}]{Observatory}
{HAWC}. 2020, {<https://www.hawc-observatory.org/> visited on 23.09.2020}

\bibitem[{Holler {et~al.}(2015)Holler, Berge, van Eldik, Lenain, Marandon,
  Murach, de~Naurois, Parsons, Prokoph, \& Zaborov}]{Holler2015}
Holler, M., Berge, D., van Eldik, C., {et~al.} 2015

\bibitem[{Kafexhiu {et~al.}(2014)Kafexhiu, Aharonian, Taylor, \&
  Vila}]{Kafexhiu2014}
Kafexhiu, E., Aharonian, F., Taylor, A.~M., \& Vila, G.~S. 2014, Physical
  Review D, 90, 123014

\bibitem[{Lipari \& Vernetto(2020)}]{Lipari2020}
Lipari, P. \& Vernetto, S. 2020, Astroparticle Physics, 120, 102441

\bibitem[{Maldera {et~al.}(2019)Maldera, Wood, Caputo, Rando, Charles, Digel,
  \& Baldini}]{Maldera2019}
Maldera, S., Wood, M., Caputo, R., {et~al.} 2019, {Fermi-LAT Performance
  <https://www.slac.stanford.edu/exp/glast/groups/canda/lat{\_}Performance.htm>
  visited on 08.09.2020}

\bibitem[{Miville-Desch{\^{e}}nes {et~al.}(2016)Miville-Desch{\^{e}}nes,
  Murray, \& Lee}]{Miville-Deschenes2016}
Miville-Desch{\^{e}}nes, M.-A., Murray, N., \& Lee, E.~J. 2016, The
  Astrophysical Journal, 834, 57

\bibitem[{Mori(2009)}]{Mori2009}
Mori, M. 2009, Astroparticle Physics, 31, 341

\bibitem[{Neronov {et~al.}(2017)Neronov, Malyshev, \& Semikoz}]{Neronov2017}
Neronov, A., Malyshev, D., \& Semikoz, D.~V. 2017, Astronomy and Astrophysics,
  606

\bibitem[{Padovani {et~al.}(2020)Padovani, Ivlev, Galli, Offner, Indriolo,
  Rodgers-Lee, Marcowith, Girichidis, Bykov, \&
  Diederik~Kruijssen}]{Padovani2020}
Padovani, M., Ivlev, A.~V., Galli, D., {et~al.} 2020, {Impact of Low-Energy
  Cosmic Rays on Star Formation}

\bibitem[{Peron {et~al.}(2021)Peron, Aharonian, Casanova, Yang, \&
  Zanin}]{Peron2021}
Peron, G., Aharonian, F., Casanova, S., Yang, R., \& Zanin, R. 2021,
  Astrophysical Journal Letters, 907

\bibitem[{Peron {et~al.}(2020)Peron, Aharonian, Casanova, Zanin, \&
  Romoli}]{Peron2020}
Peron, G., Aharonian, F., Casanova, S., Zanin, R., \& Romoli, C. 2020, The
  Astrophysical Journal Letters, 896

\bibitem[{Pothast {et~al.}(2018)Pothast, Gaggero, Storm, \&
  Weniger}]{Pothast_2018}
Pothast, M., Gaggero, D., Storm, E., \& Weniger, C. 2018, Journal of Cosmology
  and Astroparticle Physics, 2018, 45

\bibitem[{Sinha {et~al.}(2021)Sinha, Baghmanyan, Peron, Gallant, Casanova,
  Holler, \& Mitchell}]{Sinha2021SearchSinha}
Sinha, H. E. S. S.~A., Baghmanyan, V., Peron, G., {et~al.} 2021, in Proceedings
  of Science(ICRC2021) 790, 12--23

\bibitem[{Stone {et~al.}(2019)Stone, Cummings, Heikkila, \& Lal}]{Stone2019}
Stone, E.~C., Cummings, A.~C., Heikkila, B.~C., \& Lal, N. 2019, Nature
  Astronomy, 3, 1013

\bibitem[{{The Fermi-LAT collaboration}(2019)}]{TheFermi-LATcollaboration2019}
{The Fermi-LAT collaboration}. 2019, The Astrophysical Journal Supplement
  Series, 242

\bibitem[{Vos \& Potgieter(2015)}]{Vos2015NEW23/24}
Vos, E.~E. \& Potgieter, M.~S. 2015, The Astrophysical Journal, 815, 119

\bibitem[{Yang {et~al.}(2016)Yang, Aharonian, \& Evoli}]{Yang2016}
Yang, R., Aharonian, F., \& Evoli, C. 2016, Physical Review D, 93

\bibitem[{Yang {et~al.}(2015)Yang, Jones, \& Aharonian}]{Yang2015}
Yang, R.-Z., Jones, D.~I., \& Aharonian, F. 2015, Astronomy and Astrophysics,
  580, 90

\end{thebibliography}

\end{document}